\documentclass[twocolumn,tighten,times]{aastex631}

\usepackage{amsmath}
\usepackage{multirow}
\usepackage[abs]{overpic}
\usepackage{breakurl}
\usepackage{hyperref}
\usepackage{textcomp}
\usepackage{fancyhdr}
\usepackage{graphicx}
\usepackage{amssymb}
\usepackage[toc,page]{appendix}
\usepackage{nicefrac}
\usepackage{svg}
\usepackage[utf8]{inputenc}
\usepackage{varwidth}
\usepackage{scrextend}
\usepackage{xcolor}
\usepackage{colortbl}
\usepackage{tablefootnote}
\usepackage{array}
\usepackage[]{graphicx}
\usepackage{xspace}

\newcommand{\SAUNAS}{\texttt{SAUNAS}}

\newcommand\asr{\ref@jnl{Adv. Space Res.}}

\newcommand{\s}{$\sim$}
\newcommand{\n}{$-$}
\newcommand{\Chandra}{Chandra}
\newcommand{\Hubble}{\emph{HST}}

\newcommand{\ciao}{\texttt{CIAO}}

\definecolor{navyblue}{rgb}{0.0, 0.0, 0.5}

\newcommand{\escmarc}{s$^{-1}$ cm$^{-2}$ arcsec$^{-2}$}
\usepackage{makecell}

\begin{document}

\title{SAUNAS II: Discovery of Cross-shaped X-ray Emission and a Rotating Circumnuclear Disk \\in the Supermassive S0 Galaxy NGC 5084}

\author[0000-0003-3249-4431]{Alejandro S. Borlaff}
\thanks{}
\affiliation{NASA Ames Research Center, Moffett Field, CA 94035, USA}
\affiliation{Bay Area Environmental Research Institute, Moffett Field, California 94035, USA}

\author{Pamela M. Marcum}
\affiliation{NASA Ames Research Center, Moffett Field, CA 94035, USA}

\author[0000-0002-8341-342X]{Pasquale Temi}
\affiliation{NASA Ames Research Center, Moffett Field, CA 94035, USA}

\author[0000-0002-1598-5995]{Nushkia Chamba}
\thanks{NASA Postdoctoral Program Fellow}
\affiliation{NASA Ames Research Center, Moffett Field, CA 94035, USA}

\author[0000-0003-0346-6722]{S. Drew Chojnowski}
\thanks{NASA Postdoctoral Program Fellow}
\affiliation{NASA Ames Research Center, Moffett Field, CA 94035, USA}

\author[0000-0001-5357-6538]{Enrique Lopez-Rodriguez}
\affiliation{Kavli Institute for Particle Astrophysics \& Cosmology (KIPAC), Stanford University, Stanford, CA 94305, USA}

\author[0000-0002-0905-7375]{Aneta Siemiginowska}
\affiliation{Harvard Smithsonian Center for Astrophysics, 60 Garden St, Cambridge, MA 02138, USA}

\author[0000-0003-1250-8314]{Seppo Laine}
\affiliation{IPAC, Mail Code 314-6, Caltech, 1200 E. California Blvd., Pasadena, CA 91125, USA}

\author[0000-0002-6610-2048]{Anton M. Koekemoer}
\affiliation{Space Telescope Science Institute, 3700 San Martin Dr., Baltimore, MD 21218, USA}

\author[0000-0001-5825-7683]{Kelly N. Sanderson}
\affiliation{Department of Astronomy, New Mexico State University, Las Cruces, NM, 88003, USA}
\affiliation{National Radio Astronomy Observatory, Domenici Science Operations Center, PO Box 0, Socorro, NM 87801, USA}

\author[0009-0003-6389-4101]{Audrey F. Dijeau}
\affiliation{Department of Astronomy, New Mexico State University, Las Cruces, NM, 88003, USA}

\author[0000-0001-8302-0565]{Moire K. M. Prescott}
\affiliation{Department of Astronomy, New Mexico State University, Las Cruces, NM, 88003, USA}

\author{Leslie Proudfit}
\affiliation{Universities Space Research Association, NASA Ames Research Center, Moffett Field, CA, 94035, USA}

\author{Michael N. Fanelli}
\affiliation{NASA Ames Research Center, Moffett Field, CA 94035, USA}



\begin{abstract}
Combining \Chandra, ALMA, EVLA, and \emph{Hubble} Space Telescope archival data  and newly acquired APO/DIS spectroscopy, we
detect a double-lobed 17~kpc X-ray emission with plumes oriented approximately perpendicular and parallel to the galactic plane of the massive lenticular galaxy NGC\,5084 at 0.3--2.0~keV.
We detect a highly inclined ($i=71.2^{+1.8\circ}_{-1.7}$), molecular circumnuclear disk ($D=304^{+10}_{-11}$ pc) in the core of the galaxy rotating (V$^{\rm (2-1) CO}_{\rm rot}=242.7^{+9.6}_{-6.4}$ km s$^{-1}$) in a direction
perpendicular to that of the galactic disk, implying a total mass of $\log_{10}\left( \frac{M_{\rm BH}}{M_{\odot}} \right) = 7.66^{+0.21}_{-0.15}$ for NGC\,5084's supermassive black hole. Archival EVLA radio observations at 6 cm and 20 cm reveal two symmetric radio lobes aligned with the galactic plane, extending to a distance of $\overline{R}=4.6\pm0.6$ kpc from the core, oriented with the polar axis of the circumnuclear disk. The spectral energy distribution lacks strong emission lines in the optical range.
Three formation scenarios are considered to explain these multi-wavelength archival observations: 1) AGN re-orientation caused by accretion of surrounding material, 2) AGN-driven hot gas outflow directed along the
galactic minor axis, or 3) a starburst / supernovae driven outflow at the core of the galaxy. This discovery is enabled by new imaging analysis tools
including \SAUNAS\ (Selective Amplification of Ultra Noisy Astronomical Signal), demonstrating the abundance of information still to be exploited in the vast and growing astronomical archives.
\end{abstract}



\section{Introduction} \label{sec:intro}
Observational evidence suggests that supermassive black holes (SMBHs; $M_{\rm BH}=10^{6-9.5}$) are present at the core of most massive galaxies \citep{kormendy+2013araa51_511}. The accretion of material by SMBHs has been observed to be tightly connected with the evolution of the entire galaxy, enabling periods of nuclear activity in which the total luminosity of the galaxy is enhanced.

The main hypothesis of the unified scheme of active galactic nuclei (AGNs) is that the broad range of spectro-photometric properties is caused by the angle of inclination between the observer and an obstructing circumnuclear disk \citep{antonucci1993araa31_473, urry+1995pasp107_803}. The circumnuclear disk is a dusty and molecular gas component of outflowing and inflowing material surrounding the AGN, the so-called `torus'. According to this theory, when AGN's are observed at low inclination angles \citep[$i<45$--$60^{\circ}$, where $i=0^{\circ}$ would be face-on,][]{marin2014mnras441_551}, broad optical and ultraviolet (UV) spectral features (full width at half maximum; FWHM $\geq 1000-20,000$ km s$^{-1}$) are expected, corresponding to the Broad Line Region (BLR); a bright, non-stellar, central compact source visible at all wave-lengths, and with low polarization degree \citep[Type 1 AGNs,][]{netzer2015araa53_365}. At higher inclination angles, the BLR would be obscured by the torus showing only the narrow line region emission (NLR, Type 2 AGNs). See \citet{peterson2006incollection_77, netzer2015araa53_365, ramosalmeida+2017nat1_679} for reviews in the field.


A large fraction of the total mass of SMBHs is thought to have been accreted during the peak of quasar activity in the early Universe \citep[$z=2-3$,][]{boyle+1998mnras293_49, delvecchio+2014mnras439_2736, madau+2014araa52_415}. During that period ($z\sim2-3$), the star-formation rate (SFR) and SMBH growth rate peaked and then decreased by a factor of ten from $z=1$ to the local Universe \citep{shankar+2009apj690_20}. The observed decrease in SFR is coincident with one of the most notable changes in the population of galaxies: the decline of the population of spirals and the rise of lenticular galaxies in the Local Universe \citep{dressler+1997apj490_577}.


Lenticular galaxies seem to acquire their morphology by moving along multiple evolutionary paths
\citep[accretion, mergers, secular evolution, thermal stripping of the gas content][]{laurikainen+2010mnras405_1089, larson+1980apj237_692, moore+1996nat379_613, barway+2009mnras394_1991, borlaff+2014aap570_103, elichemoral+2018aap617_113, frasermckelvie+2018mnras481_5580}. However, the identification of the formation mechanism for particular galaxies continues to be a difficult task. Observational works have found signs of past \citep{machacek+2010apj711_1316} and ongoing \citep{wang+2019apj870_132} accretion events in the extended X-ray emission of lenticular galaxies, associated to the hot gas phase of the ISM, as well as in their core morphology \citep{juravnova+2019mnras484_2886}. In addition, multiple studies point towards the action of supernovae feedback, AGN activity, and their associated galactic winds as the cause for the soft X-ray band emission. \par

\begin{figure*}[t!]
\begin{center}
\includegraphics[trim={0 20 0 300}, clip, width=\textwidth]{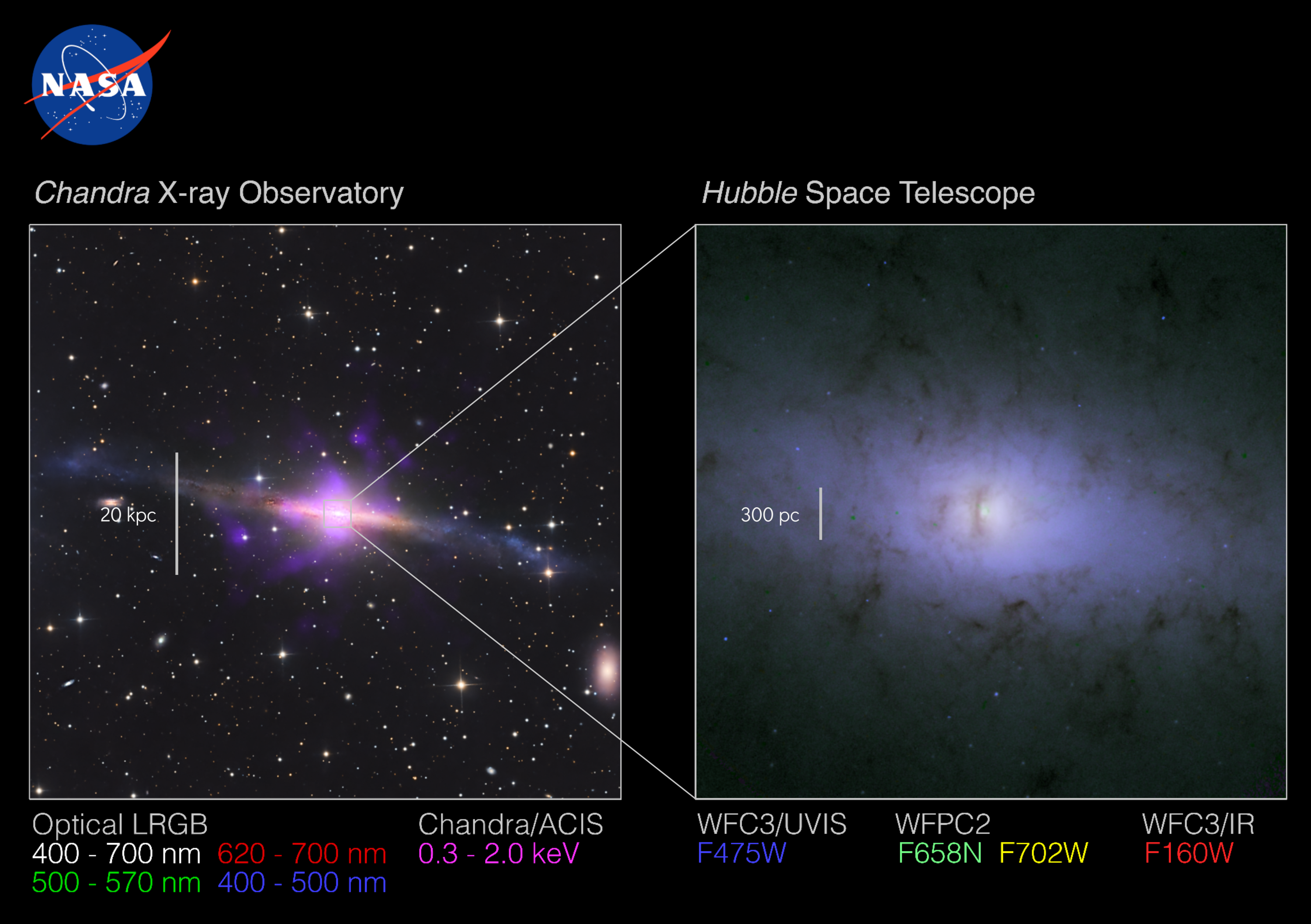}
\caption{Morphological structure of NGC\,5084 in X-ray, optical, and near-infrared wavelengths. \emph{Left panel:} Optical and X-ray image of
the disk of NGC\,5084. \emph{Background:} Luminance-RGB image (Telescope: CDK17. Camera: SBIG STXL-11002M. Filter: Astrodon Gen 1 LRGB E-Series. Credit: Martin Pugh \& Brian Diaz). Field-of-view (FOV): $15\times15$ arcmin$^{2}$. \emph{Right panel:} High resolution luminance-RGB image zoom of the core of NGC\,5084 based on \emph{Hubble} Space Telescope observations. \emph{Blue:} WFC3/UVIS F475W. \emph{Green:} WFPC2 F658N. \emph{Yellow:} WFPC2 F702W. \emph{Red:} WFC3/IR F160W.  FOV: $24\times24$ arcsec$^{2}$. See the legend for color-band description and the physical scale-bars in the panels for reference.}
\label{fig:NGC5084_poster}
\end{center}
\end{figure*}

In this paper, we present previously undetected features likely related to the presence of an AGN in the edge-on supermassive lenticular galaxy NGC\,5084 \citep[$\alpha=200\fdg0705$, $\delta=-21\fdg8276$ ICRS, $D=	29.9\pm2.1$~Mpc, $z=0.005741$, 6.90~arcsec~kpc$^{-1}$,][Fig.\,\ref{fig:NGC5084_poster}]{gottesman+1986mnras219_759, devaucouleurs+1991book, koribalski+2004aj128_16} using optical and near-infrared (NIR) Hubble, Chandra X-ray, EVLA radio continuum, and ALMA observations.
NGC\,5084 is one of the most massive lenticular galaxies in the Local Universe \citep[]{ohlson+2024aj167_31}, with a total dynamical mass of approximately $1.3\times10^{12}$~M$_{\odot}$ \citep{koribalski+2004aj128_16} and a mass-to-light ratio $\Upsilon \geq 65$. Several features of this galaxy make it an interesting case study. \par

NGC\,5084 hosts signs of past interactions which are visible in the outer regions of the galactic disk, such as a warped disk at low surface brightness intensity levels \citep{zeilinger+1990mnras246_324}, potentially related to the population of satellite galaxies within its group \citep{carignan+1997aj113_1585}. NGC\,5084 shows an unusual ratio between the $B$-band optical radius $R_{25}$ and the $K_S$-band
effective radius $R_{\rm e}$. For this galaxy, $R_{25}$ is 20 times larger than $R_e$, in contrast to $R_{25}=3.6 \times R_e$, the nominal average for spirals and S0s measured by  \citet[][]{williams+2009mnras400_1665}.  \citet{zaw+2019apj872_134} classified NGC\,5084 as an AGN according to the \citet{kewley+2001apj556_121} optical spectral line ratio criteria, based on 6dF Galaxy Survey spectra \citep{jones+2004mnras355_747,jones+2009mnras399_683}, as did \citet{irwin+2019aj158_21}, based on its radio emission (see Sec.\,\ref{subsec:data_radiopol}). In addition, NGC\,5084 is the host of a large spheroidal bulge that generates an anti-truncated surface brightness profile \citep{comeron+2012apj759_98}, a potential sign of past gravitational interactions, such as minor and major mergers \citep{younger+2007apj670_269, borlaff+2014aap570_103}.

\citet{osullivan+2017mnras472_1482} analyzed the \Chandra\ X-ray observations using the Advanced CCD Imaging Spectrometer (ACIS), concluding that the X-ray emission of NGC\,5084 is primarily generated by the central core\footnote{From \citet{osullivan+2017mnras472_1482}, about NGC\,5084: \emph{``show[ing] no evidence of a hot IGM, but does detect powerlaw emission extending $\sim100$" ($\sim$10 kpc), with a central point
source contributing the majority of the emission.}"}. No further attempt to deconvolve the Point Spread Function (PSF) or to identify the source of the emission has been published until now.  NGC\,5084 is a particularly interesting target to look for evidence of past interactions with multi-wavelength (high-resolution optical combined with wide FOV X-ray imaging) archival observations from Chandra and Hubble, in addition to millimeter (mm) and submm observations (ALMA) and radio continuum observations (EVLA).

This project is the second publication in a series that will study the hot gas halos around galaxies using X-ray observations from the \Chandra\ X-ray observatory. The first paper \citep[][\texttt{SAUNAS I}, hereafter]{borlaff+2024apj967_169} describes the \SAUNAS\ (Selective Amplification of Ultra Noisy Astronomical Signal) pipeline to detect low surface brightness emission in \Chandra/ACIS observations. See \texttt{SAUNAS I} for the details and description of the X-ray image reduction methodology. \par
This paper is organized as follows. The datasets and methodology pipeline is described in Sec.\,\ref{sec:methods}. The results are presented in Sec.\,\ref{sec:results}. The discussion and conclusions are presented in Sec.\,\ref{sec:DIS} and \ref{sec:CON}, respectively. All magnitudes are in the AB system \citep{oke1971apj170_193} unless otherwise noted.

\section{Methods and datasets} \label{sec:methods}

\begin{deluxetable*}{cccccc}
\tabletypesize{\footnotesize}
\tablecolumns{7}
\tablewidth{0pt}
\vspace{-0cm}
\tablecaption{\Chandra, \Hubble, and ALMA archival datasets \label{tab:Observations}}
\tablehead{
 \multicolumn{2}{l}{\thead[t]{\textbf{a) \Chandra\ X-ray Observatory}}}& \thead[t]{} & \colhead{} & \thead[t]{} & \colhead{}\\
\hline
\thead[t]{Obs. ID} & \thead[t]{Instrument} & \thead[t]{Exposure time} & \colhead{Mode} & \thead[t]{Count rate} & \thead[t]{Obs. date}\\
\colhead{(1)} & \colhead{(2)} & \colhead{(3)} & \colhead{(4)} & \colhead{(5)} & \colhead{(6)}\\
\colhead{} & \colhead{} & \colhead{[$ks$]} & \colhead{} & \colhead{[s$^{-1}$]} & \colhead{}}
\vspace{0.5cm}
\startdata
12173 & ACIS-I & 9.92 & VFAINT & 3.22 & 2011/08\\\hline
\vspace{0.2cm}\\
\hline\hline \multicolumn{2}{l}{\thead[t]{b) \textbf{\emph{Hubble} Space Telescope}}} & \thead[t]{} & \colhead{} & \thead[t]{} & \colhead{}\\\hline
\thead[t]{Obs. ID} & \thead[t]{Instrument} & \thead[t]{Exposure time} & \colhead{Filter} & \thead[t]{$\lambda$} & \thead[t]{Obs. date}\\
(1) & (2) & (3) & (4) & (5) & (6)\\
 &  & [$s$] & & [$\mu$m] & \\\hline
6785 & WFPC2 & 1200 & F658N &
0.658-0.660 & 1996/06 \\
6785 & WFPC2 & 500 & F702W &
0.61-0.75 &
1996/06 \\
15909 & WFC3/IR & 1197 & F160W & 1.39-1.69 & 2020/05 \\
15909 & WFC3/UVIS & 602 & F475W & 0.397-0.554& 2020/05 \\
\hline\hline \multicolumn{2}{c}{\thead[t]{c) \textbf{ALMA Radio Telescope}}} & \thead[t]{} & \colhead{} & \thead[t]{} & \colhead{}\\\hline
\thead[t]{Obs. ID} & \thead[t]{Array} & \thead[t]{Exposure time} & \colhead{Band} & \thead[t]{$\nu$} & \thead[t]{Obs. date}\\
(1) & (2) & (3) & (4) & (5) & (6)\\
 &  & [$s$] & & [GHz] & \\\hline
{2015.1.00598.S} & 12m & 544 & 6 &
228.3-248.1 & 2016/04\\
\enddata
\tablecomments{\Chandra, \Hubble, and ALMA archival datasets analyzed in this work. EVLA processed radio maps are available at the CHANG-ES project webpage: \url{https://projects.canfar.net/changes/ngc-5084/}. \emph{a) Top table}: \Chandra/ACIS observations available within 10~arcmins of NGC\,5084, retrieved from the \Chandra\ Data Archive, as of January 2024. \emph{b) Middle table}: \Hubble\ observations that include the core of NGC\,5084, retrieved from MAST, as of January 2024.  \emph{b) Bottom table}: ALMA observations of NGC\,5084 retrieved from ALMA Science Portal. Col.(1) Observation ID; Col.(2) instrument or configuration; Col.(3) total exposure time per observation; Col.(4) observing mode, filter or band; Col.(5) average count rate, wavelength or frequency range; Col.(6) exposure start date.}
\end{deluxetable*}

\subsection{Chandra X-ray Observatory} \label{sec:data_chandra}

\Chandra/ACIS is particularly well-suited to detect hot gas emission in galaxies, due to its high-sensitivity in the 0.3-2.0 keV range and the high spatial resolution that allows masking of contamination from point sources. \Chandra/ACIS observations are reduced following the \SAUNAS\ pipeline, a methodology presented in \texttt{SAUNAS I}, which specializes in the detection of low surface brightness emission from \Chandra/ACIS observations. \SAUNAS\ comprises several steps for calibration including: 1) standard pre-calibration using \texttt{CIAO}, 2) automatic point spread function modeling and deconvolution, 3) estimation of the uncertainties via bootstrapping and Monte Carlo simulations, and 4) adaptive Voronoi binning. \par
The two main products of \SAUNAS\ are surface brightness and signal-to-noise ratio maps, allowing the observer to identify potential sources and their extension up to a certain statistical limit. Since the objective is to identify the underlying shape of the X-ray extended emission in NGC\,5084, PSF deconvolution is a particularly critical step. \citet{borlaff+2024apj967_169} provides a full description of the methodology.

A total of 9.92~ks of \Chandra\ observations, using the \texttt{VFAINT} mode, have been archived for NGC\,5084 (ACIS-I, Obs.\,ID:12173, PI:~Stephen~Murray; August~2011 under \Chandra\ Cycle~12, see Table \ref{tab:Observations}). We analyze NGC\,5084's X-ray emission in four different bands: 0.3 -- 1.0~keV (soft), 1.0 -- 2.0~keV (medium), 0.3 -- 2.0~keV (broad), and 2.0 -- 8.0~keV (hard). PSF models were generated for each band, taking into account the spectra of the source, following the prescriptions from \texttt{MARX}\footnote{\url{https://cxc.cfa.harvard.edu/ciao/threads/marx_sim/}}. After PSF deconvolution, point sources -- potentially associated with background objects, X-ray binaries, or AGNs -- are identified and removed from the resulting images. Finally, the resulting frames are adaptive smoothed, and the background is subtracted.

\SAUNAS\ is designed for the detection of very diffuse, low surface brightness emission in \Chandra/ACIS observations (i.e., hot gas), as opposed to the detection of point-sources (i.e., AGNs), the latter contribution of which is removed from the final images. While the point source (FWHM$\sim1.1"$) sensitivity of \Chandra/ACIS is $\sim4\times10^{-15}$ erg cm$^{-2}$ s$^{-1}$ in 10$^4$ s, (0.4--6.0 keV), spatial binning over large scales yields deeper sensitivities.

Adaptive spatial binning techniques (Voronoi, CSMOOTH) \citep{cappellari+2003mnras342_345, ebeling+2006mnras368_65} have been widely applied in X-ray astronomy for several decades \citep[see][and references therein]{ebeling+2007apj661_33,gonzalezmartin+2009aap506_1107,broos+2010apj714_1582, ebeling+2010mnras407_83, xue+2011apj195_10, hodgeskluck+2012apj746_167, wang+2024apj962_188} allowing the detection of extended sources below the canonical point source sensitivity of the observations. While \texttt{SAUNAS I} \citep{borlaff+2024apj967_169} is dedicated to the presentation, benchmarking, and testing of the adaptive binning technique used by the pipeline, Appendix \ref{Appendix:xray_tests} in this paper provides a more in-depth discussion of point source and extended source sensitivities, and a series of tests to further verify the significance of the X-ray emission of NGC\,5084 reported in Sec.\,\ref{subsec:results_xray_ima}.

\subsection{Hubble Space Telescope Imaging} \label{sec:data_hst}

\emph{Hubble} Space Telescope observations of NGC\,5084 are available at the Mikulski Archive for Space Telescopes\footnote{\url{https://mast.stsci.edu/portal/Mashup/Clients/Mast/Portal.html}} (MAST, \dataset[doi: 10.17909/j04e-sq50]{https://doi.org/10.17909/j04e-sq50}). In particular, we focus on high-resolution imaging observations obtained with WFPC2 (Proposal PI: Malkan, Matthew A.,
Proposal ID: 6785, F702W and F658N, June 1996), WFC3/IR (F160W, Proposal PI: Boizelle, Benjamin, Proposal ID: 15909, and WFC3/UVIS (F475W, same Proposal ID as WFC3/IR, see Table \ref{tab:Observations}). Planetary Camera (PC) observations of WFPC2 allow for an angular resolution of $0.05$ arcsec, while WFC3s IR and UVIS channels have a resolution of $0.13$ and $0.04$ arcsec respectively. At a distance of $D=	29.91\pm2.12$~Mpc \citep[6.90~arcsec~kpc$^{-1}$,][]{koribalski+2004aj128_16}, and assuming a Nyquist sampled PSF, the physical spatial resolution scales are $\sim12$ pc (WFC3/UVIS) and $\sim36$ pc (WFC3/IR), respectively. Table\,\ref{tab:Observations} summarizes the available observations.

All observations were retrieved from MAST as level~3, after photometric and astrometric calibration, combination (drizzling) of the different exposures when available. The pixel scale for the final HST mosaics is 0.05~arcsec~px$^{-1}$. We detected a shift between the astrometric solution in the different filters (less than 0.5 arcsec). The offset is likely due to different guide-stars used for the various HST observations, combined with the uncertainties in the guide-star catalog. Due to high intensity of the central source and the reduced field of view of the Hubble frames, there are no sufficient reference stars or background sources to perform an automatic astrometric calibration the images. The images were therefore realigned by a simple translation of the reference pixel in the WCS, allowing the position of maximum flux to be recentered in the image (the galactic core). All analysis involving \Hubble\ images was performed on the individual filter frames (with exception of the color maps in Fig.\,\ref{fig:NGC5084_color}), and as a consequence their relative alignment does not impact the results of the present work.

\subsection{ALMA observations} \label{subsec:data_ALMA}

The $J = 2 \rightarrow 1$ rotational transition of carbon monoxide and its emission line at $\nu_{\rm rest} = 230.538$ GHz, CO(2-1) hereafter, is a tracer of the dense molecular gas in the interstellar medium (ISM). Millimeter and submillimeter wavelength molecular gas observations have been used to trace the emission in the compact circumnuclear disks present in the core of AGNs \citep{garciaburillo+2021aap652_98}, whose dynamics are dominated by the mass of the SMBHs presumably present at the core, allowing the measurement of their direct gravitational mass \citep{davis+2013nat494_328,onishi+2015apj806_39} measurements. Using a sample of 19 nearby ($D<28$ Mpc) Seyfert galaxies, the dusty and molecular torus has been measured to have a median diameter of $\sim42$ pc and a molecular mass of $\sim6\times10^{5}$ M$_{\odot}$ \citep{garciaburillo+2021aap652_98}. In the local Universe, the CO (2-1) emission line is detectable in a favorable atmospheric transmission window for ground-based radio telescopes like the Atacama Large Millimeter Array (ALMA) Radio Telescope \citep{leroy+2021apj257_43}, allowing for extremely high sensitivities and spatial resolution.

NGC\,5084 was observed with ALMA as part of the WISDOM project\footnote{WISDOM: \url{https://wisdom-project.org/}} \citep[Project code: 2015.1.00598.S, PI: Martin Bureau,][]{onishi+2017mnras468_4663}. A total of 544.32\,s worth of data in Band 6 (228.273 - 248.12 GHz) were obtained, with a spatial resolution of 0.508 arcsec (83 pc), and a velocity resolution of 1.272 km s$^{-1}$. The calibrated exposures were downloaded from the NRAO ALMA Science Archive Data Portal\footnote{ALMA Science Archive: \url{https://almascience.nrao.edu/aq/}}, where two versions of the reduced data cubes are available (ALMA and the Additional Representative Images for Legacy, or ARI-L\footnote{ARI-L: \url{https://almascience.eso.org/alma-data/aril}}). \par

The results of the analysis are presented in Sec.\,\ref{subsec:results_alma}. These are compatible regardless of the version of the cubes analyzed. The preliminary inspection by the automatic detection pipelines of the ALMA Science Archive found no emission lines in any of the datacubes. In fact, none of the papers from the WISDOM project has published any result from the NGC\,5084 observations analyzed in the present work \citep{smith+2021mnras500_1933, davis+2022mnras512_1522}.

\subsection{CHANG-ES EVLA radio observations} \label{subsec:data_radiopol}

Radio-continuum observations trace the synchrotron emission that arises from the warm and diffuse interstellar medium \citep{beck+2013inbook_641}. Radio emission from AGN jets is often visible in 1.4 and 5 GHz observations \citep[20 cm and 6 cm]{sebastian+2019apj883_189, hardcastle+2020na88_101539}. NGC\,5084 is part of a sample from the Continuum Halos in Nearby Galaxies—an EVLA Survey \citep[CHANG-ES][]{irwin+2012aj144_43}. L (20 cm) and C-band (6 cm) radio observations in the B, C, and D configurations are available as part of the CHANG-ES data archive\footnote{CHANG-ES: \url{https://projects.canfar.net/changes/ngc-5084/}} \citep{wiegert+2015aj150_81, irwin+2019aj158_21}.

The CHANG-ES data products include total intensity, linear polarization degree, and polarization angle observations, which allow for a measure of the HI column density and orientation of the magnetic field (B-field). Radio polarization is a valuable tool for tracing radio jets \citep{pasetto+2018aap613_74, sebastian+2020mnras499_334}.
The analysis of the radio polarimetric observations in NGC\,5084 revealed a very low polarization degree ($p\sim0.3$\%) in the core, potentially caused by beam depolarization \citep{haverkorn+2004aap421_1011, pasetto+2018aap613_74}, and as a consequence we will not consider them for the present study. \par

\subsection{Optical spectroscopic observations} \label{subsec:data_optical_spectra}

In addition to the previous datasets, optical spectral observations can provide valuable information about the source of interest. In particular, the presence and intensity of any optical emission lines (hydrogen Balmer series lines) can reveal signs of on-going star formation (H$\alpha$), AGN activity (H$\alpha$/[NII] ratio), or their absorption by younger ($t\lesssim 2-4$ Gyr) population of stars.

NGC\,5084 presents 6dF Galaxy Survey spectra \citep{jones+2004mnras355_747,jones+2009mnras399_683} obtained with a single fiber at the core of the galaxy. In order to disentangle the potential AGN emission lines from the spectral properties of the stellar population at the core, we obtained 1200 s of slit-spectra observations using the Dual Imaging Spectrograph (DIS) at the 3.5m-Apache Point Observatory, divided into two orientations (along the major axis of the galaxy and the North-South direction, corresponding with the circumnuclear core) and the blue (B1200, $\lambda$ = [3560-4830] \AA, $0.62$ \AA\ px$^{-1}$) and red (R1200, $\lambda = [5600-6790]$ \AA, $0.58$ \AA\ px$^{-1}$) channels. The slit aperture was set to 1.5 arcsec, centered around the core of the galaxy in both orientations and channels. The results of the 6dF and APO/DIS optical spectra are presented and analyzed in Sec.\,\ref{subsec:results_Optical_spectra}.

\section{Results} \label{sec:results}

\subsection{X-ray imaging} \label{subsec:results_xray_ima}
\begin{figure*}[t!]
\begin{center}
\begin{overpic}[trim={75 0 0 40}, clip, width=\textwidth]
{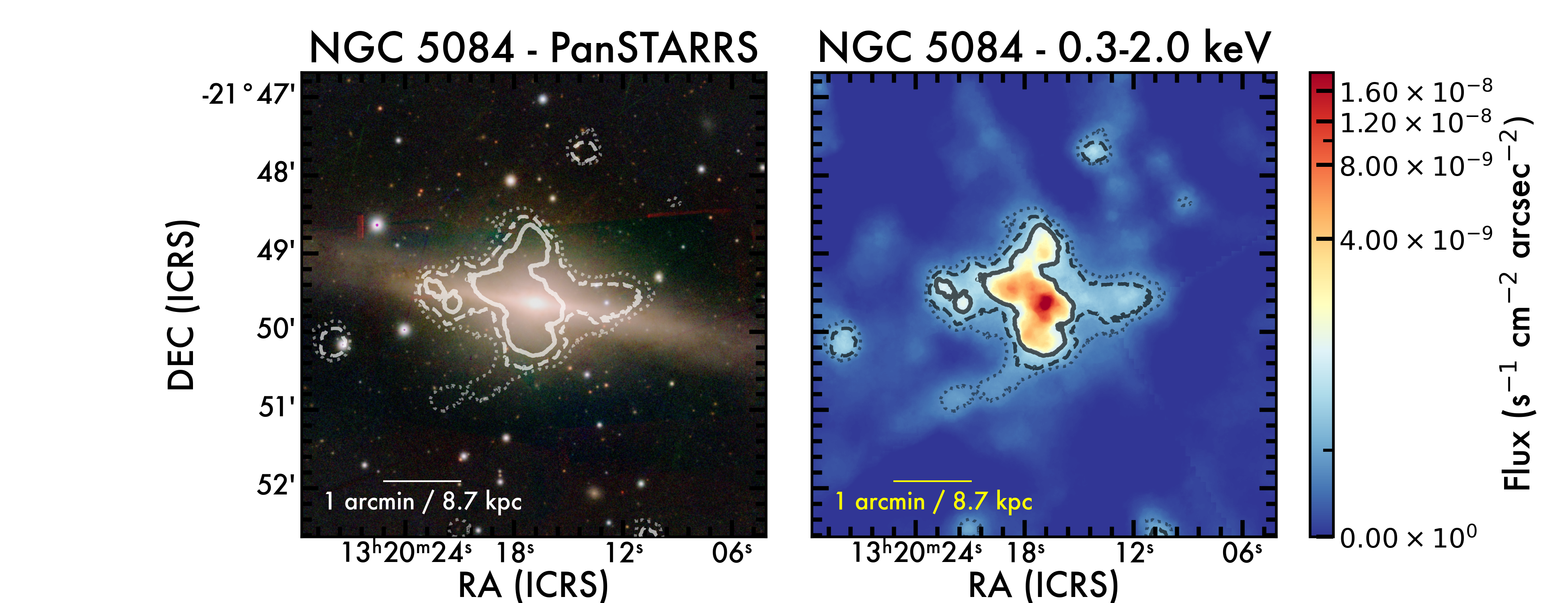}
\put(70,172.5){\Large \color{yellow} \textsf{Optical (gri)}}
\put(255,172.5){\Large \color{yellow} \textsf{0.3-2.0 keV}}
\end{overpic}
 \vspace{-0.5cm}
\caption{Diffuse X-ray emission of NGC\,5084 as detected with \SAUNAS/\Chandra\ in 0.3-2.0 keV broad band. \emph{Left:} X-ray contours (this work) over plotted on the optical $gri$ color Pan-STARRS image \citep{chambers+2016arXiv1612.05560}. \emph{Right:} \SAUNAS\ map of the diffuse X-ray emission, corrected for PSF, point-sources, and sky-background. Solid, dashed, and dotted contours represent the 5, 3, and $2\sigma$ detection levels of X-ray emission, represented in white (left panel) and black (right panel) for contrast.}
\label{fig:NGC5084}
\end{center}
\end{figure*}

Figure \,\ref{fig:NGC5084} presents the new \SAUNAS-processed images in the selected broad X-ray band (0.3--2.0 keV), overlayed on the optical morphology of the galaxy \citep[large FOV optical and near-infrared $gri$ image from Pan-STARRS, ][]{chambers+2016arXiv1612.05560}. Additionally, Fig.\,\ref{fig:NGC5084_per_band} in Appendix \ref{Appendix:Xray_subbands} shows the three X-ray (\emph{soft:} 0.3-1.0 keV, \emph{medium:} 1.0-2.0 keV, \emph{hard:} 2.0 - 8.0 keV) bands. \par

Based on the 0.3-2.0 keV emission map, NGC\,5084 shows a complex morphology in X-ray with a distinct cross-shape component, including 1) a diffuse emission perpendicular (north-south) to the disk, detectable at a $5\sigma$ level and 2) a secondary, fainter but significant ($3\sigma$), component parallel to the disk of the host galaxy (east-west). These features will be referred to in the text as the vertical and horizontal components, respectively. Between $10"<R<30"$ ($\sigma>5$) from the galactic core, the X-ray emission is elongated towards the south and north-east directions. The $3\sigma$ vertical emission extends over $\sim2$ arcmin in the north-south direction or 8.7 kpc above and below the core (a total of 17~kpc from edge to edge) while the east--west emission reaches a total extension of 2.6 arcmin (22 kpc). This result contrasts with the previous analysis of \citet{osullivan+2017mnras472_1482}, who detect the bright core and noted \emph{potential} but unconfirmed extended emission.

We analyze the total integrated flux in seven different morphological areas: (1) Total 
flux inside the $3\sigma$ contour in Fig.\,\ref{fig:NGC5084}; (2) Core ($R<10"$); (3) Extended emission ($10"<R<1.5'$, and $>3\sigma$); and (4-7) North, West, South, and East lobes. The lobe regions are defined by dividing the extended emission in 4 quadrants (with separations at 45$^{\circ}$, 135$^{\circ}$, 225$^{\circ}$, 315$^{\circ}$), excluding the emission from the core ($R<10"$) with a limit at $R=1.5'$. The separation between the core and extended regions $R=10"$ is defined based on the inspection of the X-ray morphology as the maximum radius where the X-ray emission does not show a significant elongation. The results are presented in Table \ref{tab:Xray_flux}. The fluxes take into account PSF deconvolution, and they exclude the emission from point sources identified by the \Chandra\ catalog of point sources (XRBs, AGN core). The analysis shows that the integrated emission from the north and east lobes is in excess of a 10$\sigma$ detection. The south and west are substantially dimmer but significant above the background at 6.6$\sigma$ and 3.2$\sigma$.

In the soft and medium sub-bands (see Appendix \ref{Appendix:Xray_subbands}, Fig.\,\ref{fig:NGC5084_per_band}) the north--south and east--west emission components are detectable. The soft band (0.3--1.0 keV) shows a mostly vertical emission, with a slightly distorted structure parallel to the main disk and extending towards the east. The medium 1.0--2.0~keV band displays a more axisymmetric morphology with significant emission in the west lobe. The hardest band (2.0--8.0 keV) shows signs of emission as unconnected blobs in the southern and east lobes at a ($2-3\sigma$) level, oriented in the same direction as detected in the broad and soft bands, while the northern and western lobes are undetected at $3\sigma$. The location of the X-ray emission centroids -- measured as the maximum surface brightness intensity peak -- is compatible between the three bands at a $1\sigma$ level.

In order to verify the results from the pipeline, we study the significance of the extended emission in Appendix \ref{Appendix:Xray_noPSFdeco_test} using two different, additional methodologies widely used in the literature. First, we determine if there is an excess of emission around the bright core of the galaxy by comparing the PSF surface brightness profile with the observed profile in the original \Chandra/ACIS observations, without applying Voronoi binning or PSF deconvolution. This methodology has been extensively used in the literature \citep{fabbiano+2017apj842_4,fabbiano+2018apj855_131, jones+2020apj891_133,ma+2020apj900_164,ma+2023apj948_61} to identify hot gas halos and other extended emission components. The results (see Fig.\,\ref{fig:NGC5084_psf_profile}) show a clear excess of emission above that expected PSF scattering up to the same radial distance as predicted by the $3\sigma$ contours on Fig.\,\ref{fig:NGC5084}, confirming that the X-ray emission of NGC\,5084 is not caused by PSF-scattered light from the bright core and that the extended emission is significant.

Second, we analyze the X-ray emission in fixed apertures over the observed lobes, by applying a Poisson test \citep{KRISHNAMOORTHY200423} to determine if the number of events is compatible with that of the background. The test rejects the null hypothesis that the emission in the apertures centered over the extended lobes is compatible with the background with $p<0.05$, for each of the lobes.

In summary, the results from three different methodologies (\SAUNAS, PSF-comparison, and Poisson tests) confirm the same conclusions: (1) the extended cross-shaped X-ray emission around NGC\,5084 is statistically significant; (2) the PSF scattered-light from the core is not sufficient to explain the extended emission; and (3) the the four X-ray lobes around NGC\,5084 are detectable at a $>3\sigma$ level.

\begin{deluxetable*}{lccccc}
\tabletypesize{\footnotesize}
\tablecolumns{7}
\tablewidth{0pt}
\tablecaption{Broad-band (0.3 - 2.0 keV) X-ray emission flux per region  \label{tab:Xray_flux}}
\tablehead{
\thead[t]{Region}& \thead[t]{Position} & \thead[t]{Area} & \thead[t]{Energy flux} & \thead[t]{Photon flux} & \thead[t]{SNR} \\
 & {[}$\alpha$, $\delta$, ICRS{]} & {[}arcmin$^2${]} & {{[} 10$^{-15}$ erg cm$^{-2}$ s$^{-1}${]}}  & {{[} 10$^{-6}$ cm$^{-2}$ s$^{-1}${]}}  & {[$\sigma$]}\\
\hline
\colhead{(1)} & \colhead{(2)} & \colhead{(3)} & \colhead{(4)} & \colhead{(5)} & \colhead{(6)}}
\vspace{0.5cm}
\startdata
All & $200.070^{\circ}$, $-21.828^{\circ}$ & 2.37 & $40.00\pm0.58$  & $21.71\pm0.32$ & 68.6 \\
Core [$R<10"] $& $200.070^{\circ}$, $-21.828^{\circ}$ & 0.09 & $7.14\pm0.57$ & $3.88\pm0.31$ & 12.4\\
Extended [$10"<R<1.5'$] & $200.070^{\circ}$, $-21.828^{\circ}$ & 2.28 & $32.85\pm0.60$ & $17.83\pm0.33$ & 54.3\\
North & $200.073^{\circ}$, $-21.816$ & 0.57 & $8.46\pm0.71$ & $4.59\pm0.39$ & 11.9 \\
West  & $200.088^{\circ}$, $-21.825$ & 0.78 & $5.1\pm1.6$ & $2.76\pm0.87$ & 3.2 \\
South & $200.071^{\circ}$, $-21.836$ & 0.42 & $7.2\pm1.1$ & $3.95\pm0.60$ &  6.6\\
East  & $200.056^{\circ}$, $-21.827$ & 0.50 & $11.8\pm0.78$ & $6.41\pm0.42$ & 15.1 \\
\hline
\enddata
\tablecomments{X-ray photometric properties of the different regions identified in NGC\,5084 using \Chandra/ACIS broad-band (0.3--2.0 keV) observations. Col.(1) Name of the region; Col.(2) Center in sky coordinates ($\alpha$, $\delta$, ICRS); Col.(3) Projected sky angular area; Col.(4) Integrated energy flux; Col.(5) Integrated photon flux; Col.(6) Signal-to-noise ratio, measured in $\sigma$ over the sky background emission.}
\end{deluxetable*}

\subsection{Hubble Space Telescope imaging} \label{subsec:results_hst}

\begin{figure*}[t!]
\begin{center}
\includegraphics[trim={0 160 0 0}, clip, width=\textwidth]{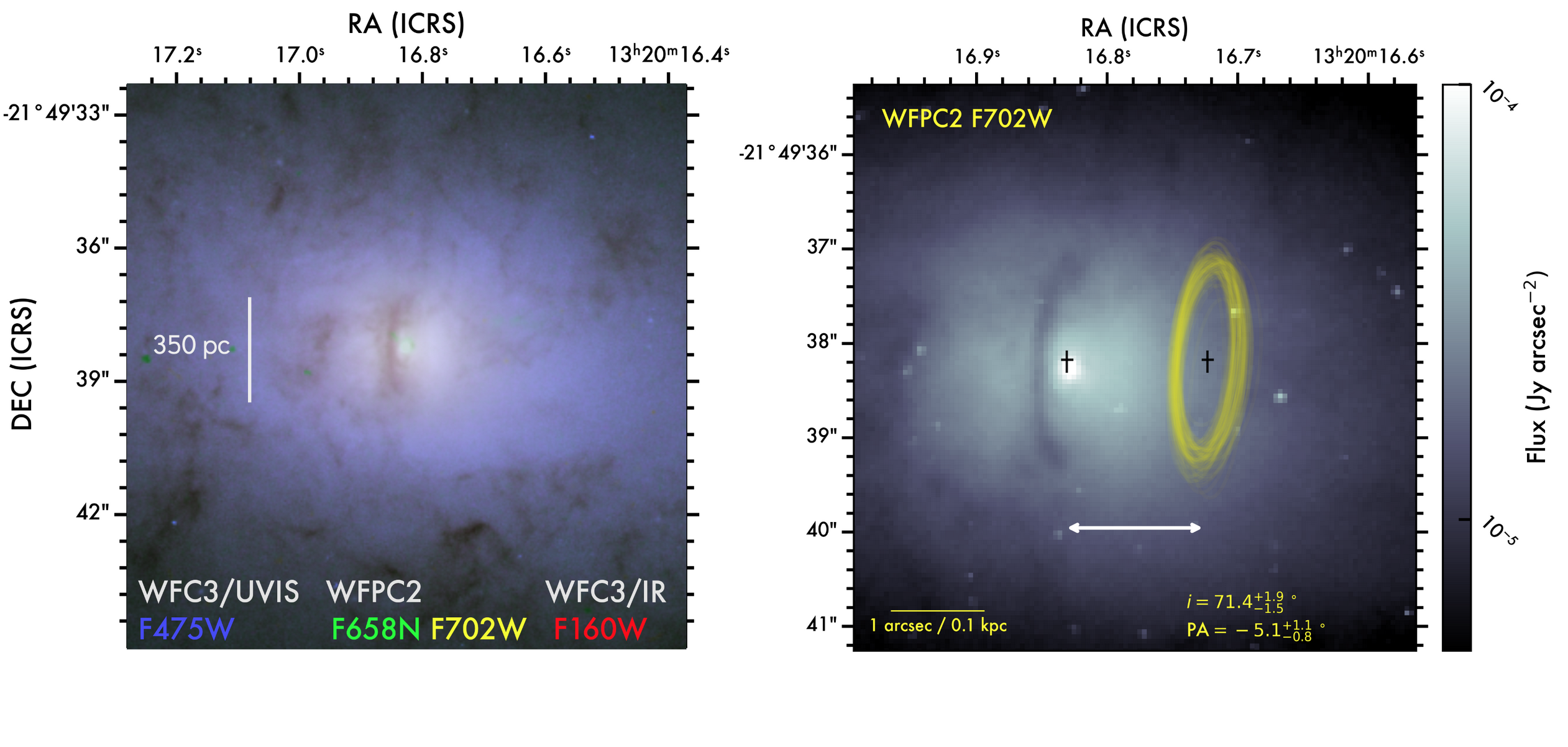}
\caption{Morphology of the core of NGC\,5084 obtained with Hubble. \emph{Left:} LRGB image generated using four HST filters: WFC3/UVIS F475W (blue), WFPC2 F658N (green), WFPC2 F702W (yellow), and WFC3/IR F160W (red). The area shown is $12.6\times12.6$~arcsec$^2$ ($1.82\times1.82$~kpc$^2$ at a distance of 29.91~Mpc). \emph{Right:} Analysis of the ellipticity of the circumnuclear disk of NGC\,5084. Yellow ellipses show the Bootstrapping + Monte Carlo elliptical fits, displaced 1 arcsec (double headed arrow shows the offset distance) to the west from the core of the galaxy to enhance the visual comparison. The results of the fit are shown in the panel. The black crosses show the best fit to the center of the fitted ellipses. HST/WFPC2 F702W flux is shown in the background. Note that the RGB color levels in the left panel have been optimized to emphasize the disk. Photometric color maps ($g-r$ and F475W - F702W) are available in Fig.\,\ref{fig:NGC5084_color}. See the scale and colorbars for reference.}
\label{fig:NGC5084_hst}
\end{center}
\end{figure*}

To further investigate the nature of this complex emission, we explored the available HST observations of NGC\,5084 (see Table \ref{tab:Observations}), presented in Fig.\,\ref{fig:NGC5084_hst}.
The image shown ($12.6\times12.6$~arcsec$^2$) corresponds to a physical area of $1.82\times1.82$~kpc$^2$ at 29.9~Mpc. The morphology of the inner core of NGC\,5084 shows a characteristic arc-shaped absorption feature perpendicular to the orientation of the galaxy's disk that partially obscures the core of the galaxy. The obscuring structure is approximately 300~pc in diameter, and it is visible in the F658N, F702W, and F475W filters. We refer to Fig.\,\ref{fig:NGC5084_multiband} in Appendix \ref{Appendix:Hubble_subbands} for the individual band images. The images reveal the presence of a substantial amount of dusty filaments around the bright core of the galaxy. The dust extinction shows a higher contrast in the shortest wavelength (F475W) observations against the continuum background and the bright unresolved core of the galaxy.

The central absorption feature is visible on the west-side of the bright core, and it presents a clear curvature centered around the core of the galaxy. The morphology of this feature is compatible with that of a highly inclined circumnuclear disk \citep{moellenhoff+1987aap174_63, jaffe+1993nat364_213,  kormendy+1994inproceedings_147, ferrarese+1996apj470_444, vandermarel+1998aj116_2220}, with a north-south angular diameter of $2.1\pm0.1$ arcsec ($D=304^{+10}_{-11}$ pc, as estimated below). By subtracting the F702W flux from the narrow band F658N filter, a proxy of the [NII, $\lambda$6548] + $H\alpha$ emission intensity map is generated. The F658N - F702W map (see Fig.\,\ref{fig:NGC5084_multiband} in Appendix \ref{Appendix:Hubble_subbands}) reveals a bright emission core surrounded by a halo, in which the dust absorption of the disk contrasts clearly.

The position angle and inclination of the circumnuclear disk were measured for comparison with those of the main components of the rest of the galaxy. To account for uncertainties, reference points across the arc dust lane using \texttt{SAODS9} were defined and then fit to an ellipse using $N=1000$ bootstrapping and Monte Carlo simulations, allowing us to estimate uncertainties. Each Monte Carlo simulation used a displaced location ($\alpha, \delta$) of the reference points within the FWHM of the PSF of the instrument. The WFPC2/F702W filter was chosen for this task due to its better angular resolution and less contamination by dust filaments in the FOV. \par

The results are shown in the right panel of Fig.\,\ref{fig:NGC5084_hst}. The model ellipse fit is shown displaced from the observed disk for clarity. The center of the ellipse is $\alpha, \delta = (200\fdg0701, -21\fdg8272) \pm 0.1$ arcsec, compatible with the location of the core of the galaxy. The disk diameter is  $D=304^{+10}_{-11}$ pc, with a median inclination of $i=71.2^{+1.8}_{-1.7}$ degrees, and a position angle of $\theta = -5.1^{+1.1}_{-1.0}$ degrees, indicating that the relative orientation of the circumnuclear disk is perpendicular to the galactic plane.

\subsection{ALMA CO(2-1) emission line observations}
\label{subsec:results_alma}

\begin{figure*}[t!]
\begin{center}
\includegraphics[trim={0 0 0 0}, clip, width=\textwidth]{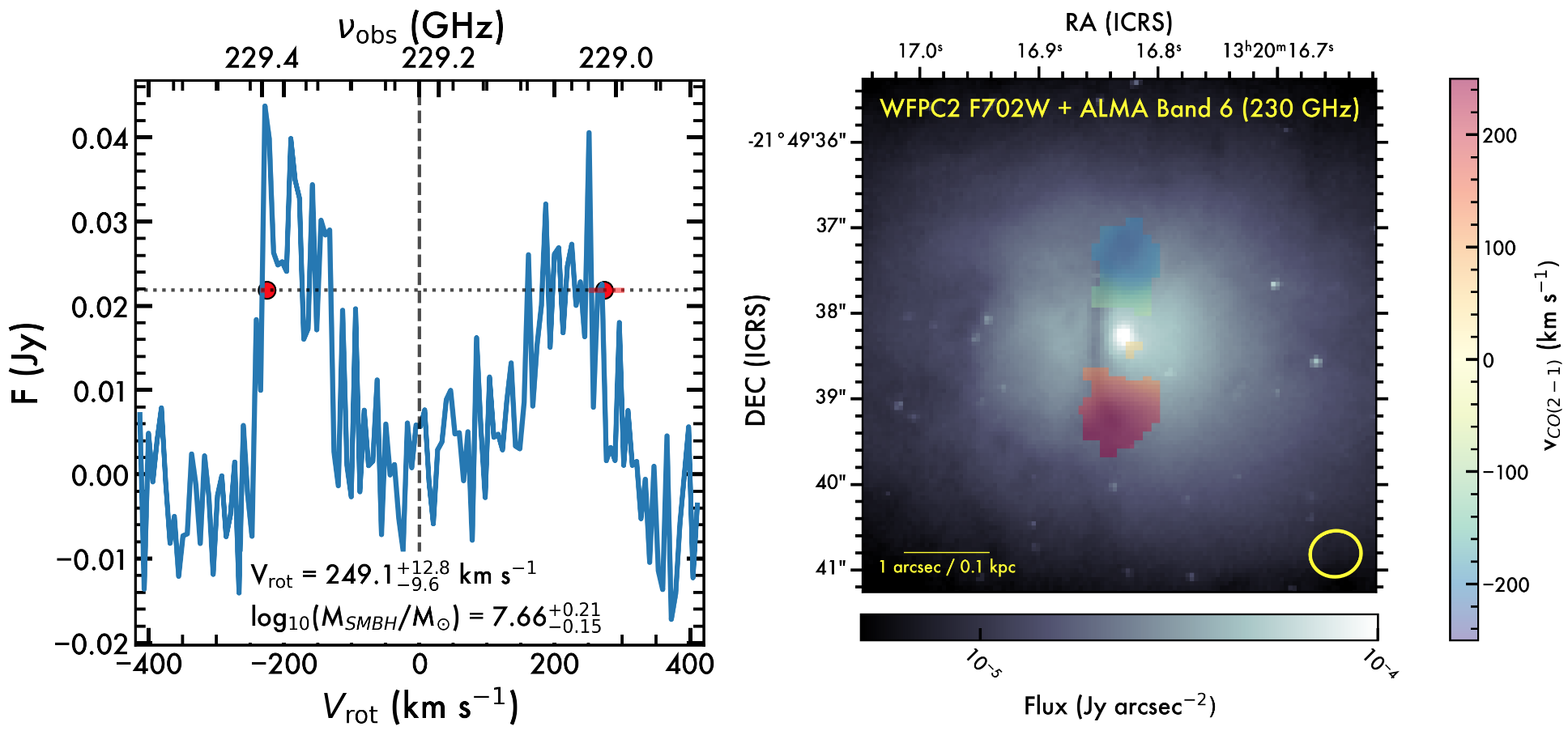}
\caption{Velocity field analysis of the circumnuclear disk of NGC\,5084, based on ALMA CO(2-1) observations. \emph{Left:} Spectral analysis of the redshifted ($z=0.005741$) CO(2-1) emission line $\nu^{\rm CO(2-1)}_{\rm rest} = 230.538$ GHz. Blue line shows the integrated spectra as a function of the frequency (top axis) and the rotation velocity (bottom). The horizontal dashed line and red dots represent the line width at half maximum. Vertical dashed line represents the center of the redshifted CO(2-1) emission line. \emph{Right:} CO(2-1) velocity map (see right colorbar) over the WFPC2/F702W intensity map (bottom colorbar). The yellow ellipse represents ALMA observations beam size ([B$_{\rm maj}$, B$_{\rm min}]= [0.60,0.54]$ arcsec).}
\label{fig:ALMA}
\end{center}
\end{figure*}

Figure \ref{fig:ALMA} shows the results from the analysis of the ALMA Band 6 spectral data cube obtained in NGC\,5084. Initial inspection of the calibrated cube did not show sufficient emission per spaxel at the original spectral resolution of 1.3 km s$^{-1}$. After rebinning the datacube in the spectral axis to a velocity resolution of 12.9 km s$^{-1}$, emission lines with SNR$>3$ are detected. We fit Gaussian models along the spectral direction for each pixel in the cube, identifying the intensity, centroid, and dispersion of the CO(2-1) emission lines (moments 0, 1, and 2).

The positions that do not contain emission with an associated intensity higher than the noise at a $95\%$ probability confidence level are masked. The velocity map (moment 1) of the CO(2-1) emission lines is represented in a color scale over the F702W HST/ACS image in the right panel of Fig.\,\ref{fig:ALMA}. The results show an unmistakable edge-on rotation pattern, with the north (south) edge of the disk moving towards (away from) the observer along the line of sight. The peaks of CO(2-1) emission are coincident with the edges of the circumnuclear disk, where the column density is expected to be highest due to projection effects.\\

The spectra of the locations with significant emission are combined (summed) to study the line profile and characterize the amplitude of the rotation pattern. The combined spectra is shown in the left panel of Fig.\,\ref{fig:ALMA}. We characterize the rotation amplitude using the line width at the half-maximum, following a similar procedure as in \citep[W$_{50}$,][]{smith+2021mnras500_1933}. The velocities (lowest and highest) at which the spectrum reaches its half-maximum are measured through interpolation of the combined spectra. This process is repeated using $N=1000$ Monte Carlo simulations, with each data point being displaced by its spectral flux uncertainty (noise level). The measured rotation velocity is V$_{\rm rot} = 249.1^{+12.8}_{-9.6}$ km s$^{-1}$ (W$_{50}=498^{+26}_{-19}$ km s$^{-1}$).\\

Finally, we use the definition in \citet{smith+2021mnras500_1933} to measure the mass of the central SMBH based on their fitted relation with the line width at the half-maximum W$_{50}$ and the disk inclination (see their Eq.\,6):
\begin{equation}
\begin{split}
\log_{10}\left( \frac{M_{\rm BH}}{M_{\odot}} \right) & = \\ (8.5\pm0.9) & \left[ \log_{10} \left( \frac{W_{50}\,{\rm [km\,s}^{-1}]}{\sin i\,\,{\rm [km\,s}^{-1}]}\right)  -2.7 \right] + (7.5\pm0.1).
\end{split}
\end{equation}
The uncertainties in their fitted relation are propagated using the Monte Carlo simulations, together with those in the inclination of the disk (see Sec.\,\ref{subsec:results_hst}), as well as the spectra. A total mass of $\log_{10}\left( \frac{M_{\rm BH}}{M_{\odot}} \right) = 7.66^{+0.21}_{-0.15}$ is determined for the SMBH of NGC\,5084. No previous estimates have been reported for the mass of the SMBH of this galaxy.

\subsection{EVLA 6 cm radio continuum observations} \label{subsec:results_radiopol}

\begin{figure}[t!]
\begin{center}
 \begin{overpic}[trim={0 50 90 0}, clip, height=7.7cm]{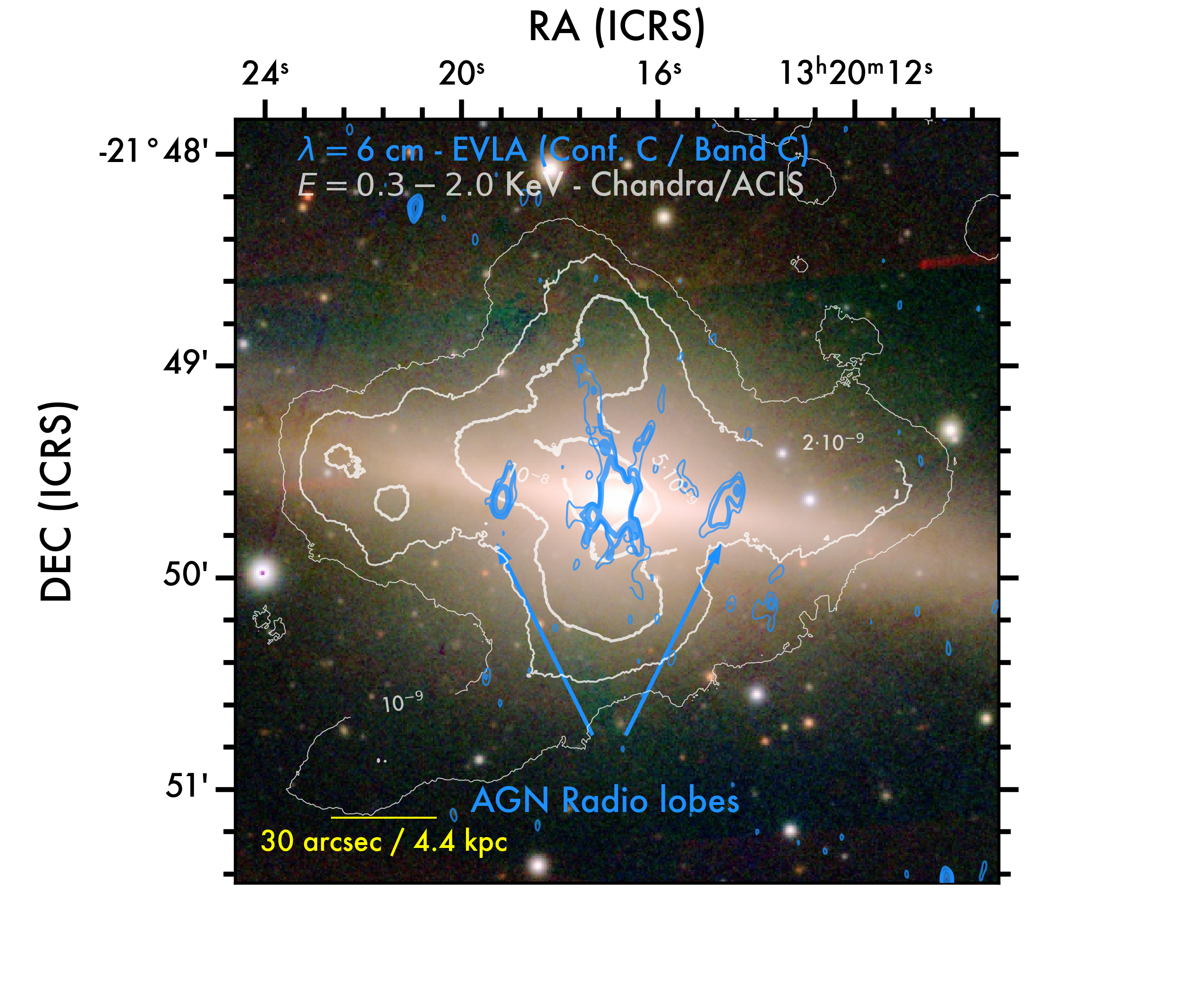}
 \end{overpic}
\caption{Comparison of EVLA radio-wavelength 6 cm observations with the optical and X-ray morphology of NGC\,5084. Blue contours represent the [3, 5, 10]$\sigma$ confidence levels at 6 cm in EVLA Band C observations from CHANG-ES \citep{wiegert+2015aj150_81}. The sensitivity for VLA C-band observations is 7.9 $\mu$Jy beam$^{-1}$ in 15.64$\times$8.35 arcsec$^{-2}$. In the background, the optical $gri$ RGB image of NGC\,5084 from the Pan-STARRS image \citep{chambers+2016arXiv1612.05560}. White contours represent the emission in the \Chandra/ACIS X-ray soft-band (0.3--2.0 keV) in \escmarc.}
\label{fig:NGC5084_EVLA}
\end{center}
\end{figure}

Fig.\,\ref{fig:NGC5084_EVLA} compares the total intensity of the EVLA 6 cm observations to the X-ray 0.3-2.0 keV using contours overlaid on a reference optical $gri$ image of NGC\,5084 and the 6 cm total surface brightness intensity. As described by \citet{wiegert+2015aj150_81}, the bright core of NGC\,5084 is dominated by systematic effects, including residual sidelobes, generating an artificial X-shape pattern that follows the beam shape. Because correcting these residuals is beyond the scope of our study, we conservatively regard the 6~cm emission in the north and south regions of NGC\,5084's core as artifacts.

Interestingly, \citet{wiegert+2015aj150_81} identifies extended emission to the east and west of the central bright core. This emission is highlighted in Fig.\,\ref{fig:NGC5084_EVLA}, and was identified as radio lobes in \citet[][see their Table 10]{irwin+2019aj158_21}. These two radio sources are detected at a 10$\sigma$ level, being located at a symmetric distance from the core. The distance from the east lobe to the core is $R=31.5^{+4.5}_{-4.5}$ arcsec or $R=4.52^{+0.66}_{-0.65}$ kpc, while the equivalent from the west lobe is $R=31.1^{+4.5}_{-4.6}$ arcsec ($R=4.51^{+0.65}_{-0.67}$ kpc), being compatible at a 3$\sigma$ confidence level. The position angles from the east and west lobes to the center are also compatible at a 3$\sigma$ level, PA$_{\rm east - core} =92.8^{+6.4\circ}_{-7.1}$, PA$_{\rm core - west} =92.2^{+7.1\circ}_{-7.6}$. Taking into account the inclination of the circumnuclear disk ($i=71.2^{+1.8\circ}_{-1.7}$), and assuming that the lobes are oriented along the line of the AGN radio jet axis, the deprojected distance to the core is $R=4.8\pm0.70$ kpc, taking into account the uncertainties in the location of the core, radio lobes, and inclination. The results are compatible at the 1.4 and 5 GHz bands. The total luminosity of each lobe in 5 GHz is $L_{\rm 5 GHz} = [1.5,1.8] \times 10^{+19}$ W Hz$^{-1}$ while the core is two orders of magnitude brighter with $L_{\rm 5 GHz, core} = 3.60\pm0.01 \times 10^{+21}$ W Hz$^{-1}$. In 1.4 GHz, the lobes are brighter $L_{\rm 1.4 GHz} = [4,6] \times 10^{+19}$ W Hz$^{-1}$, with $L_{\rm 5 GHz, core} = 3.60\pm0.01 \times 10^{+21}$ W Hz$^{-1}$ in the core. The distances to the core of each lobe and the luminosities of each component are included in Table \ref{tab:Radio_Lobes}.

\begin{deluxetable*}{ccccc}
\tabletypesize{\footnotesize}
\tablecolumns{5}
\tablewidth{0pt}
\tablecaption{Radio lobe galactocentric distance and flux   \label{tab:Radio_Lobes}}
\tablehead{
\thead[t]{Parameter} & \thead[t]{Band} & \thead[t]{Core} & \thead[t]{West} & \colhead{East}\\
\colhead{(1)} & \colhead{(2)} & \colhead{(3)} & \colhead{(4)} & \colhead{(5)}\\
\colhead{} & \colhead{} &  \colhead{} &  \colhead{} & \colhead{}}
\vspace{0.5cm}
\startdata
\multirow{2}{*}{R$_{\rm core}$ [kpc]} & 5 GHz & - & $4.51^{+0.68}_{-0.67}$  & $4.52^{+0.62}_{-0.70}$ \\
                                      & 1.4 GHz & - & $4.34^{+0.84}_{-0.90}$ & $4.32^{+0.88}_{-0.81}$ \\
\multirow{2}{*}{R$_{\rm core, depro}$ [kpc]} & 5 GHz & - & $4.77^{+0.71}_{-0.70}$  & $4.76^{+0.69}_{-0.72}$ \\
 & 1.4 GHz & - & $4.59^{+0.90}_{-0.95}$ & $4.56^{+0.93}_{-0.87}$\\
\multirow{2}{*}{L [W Hz$^{-1}$]} & 5 GHz & $3.60\pm0.04 \times 10^{+21}$ & $1.52\pm0.06 \times 10^{+19}$ &  $1.86\pm0.05 \times 10^{+19}$\\
 & 1.4 GHz & $3.60\pm0.01 \times 10^{+21}$ & $4.33\pm0.22 \times 10^{+19}$ & $6.31\pm0.16 \times 10^{+19}$\\
\hline
\enddata
\tablecomments{Photometric and morphological parameters from the radio-lobes detected in the core of NGC\,5084 using EVLA~C (5~GHz) and L-band (1.4~GHz) observations. Col.(1)  R$_{\rm core}$ is projected distance to the core in kpc, R$_{\rm core, depro}$ is inclination deprojected distance to the core in kpc, and L is total luminosity in W Hz$^{-1}$; Col.(2) Observation frequency; Col.(3) Values associated with the core; Col.(4-5) west and east lobes.}
\end{deluxetable*}


In summary, following the classification criteria in \citet{bridle+1984araa22_319}, based on the elongation, symmetry to a bright core, and deblending from other sources, we identify this emission as a radio jet emitted from the central AGN pointing towards the disk of NGC\,5084, in agreement with the classification by \citet{irwin+2019aj158_21}. We will discuss the implications of its tilted orientation in Sec.\,\ref{subsec:discussion_merger}.

\subsection{Optical spectra} \label{subsec:results_Optical_spectra}

\begin{figure*}[t!]
\begin{center}
 \begin{overpic}[trim={0 0 0 0}, clip, width=\textwidth]{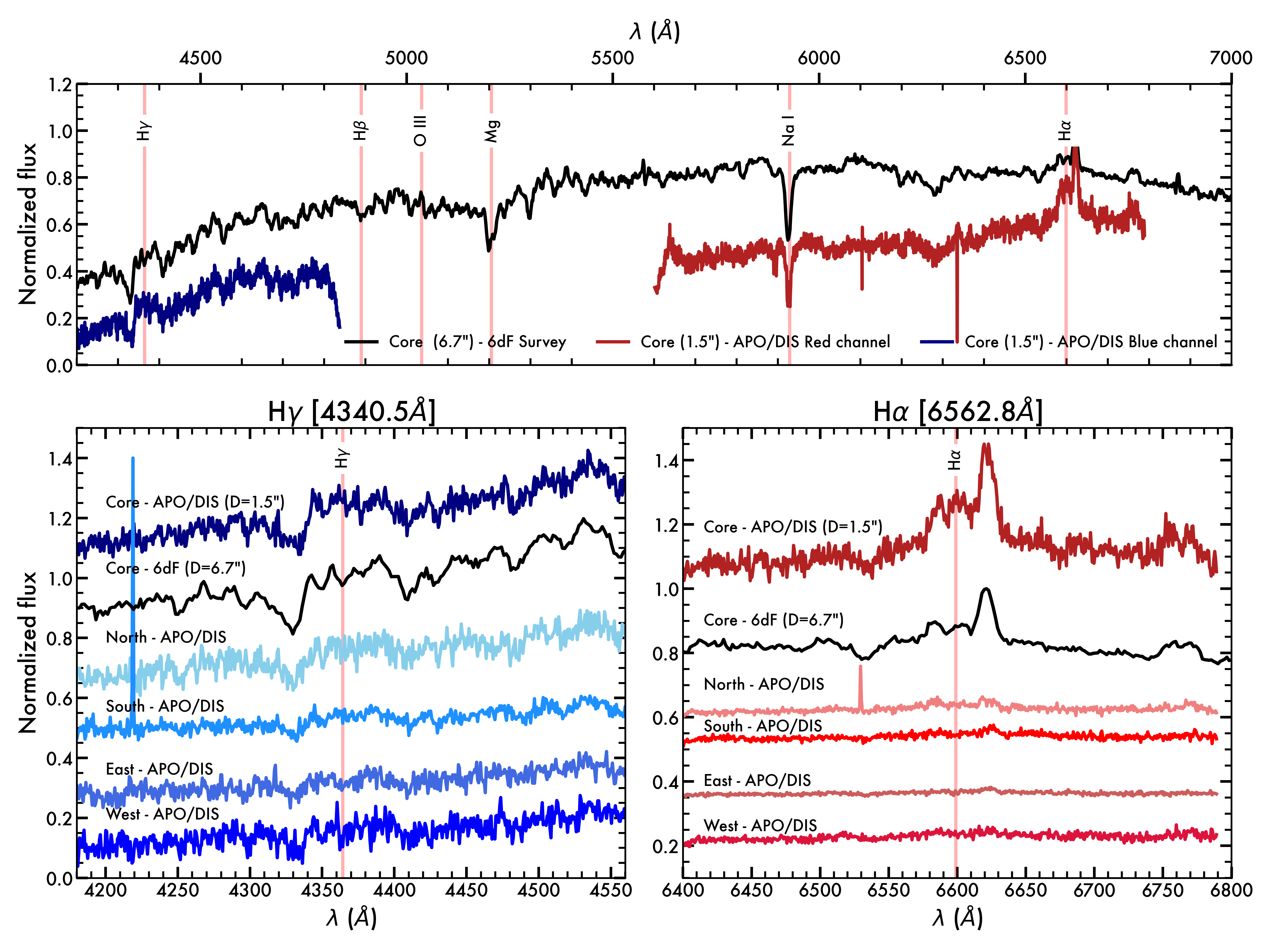}
\end{overpic}
\vspace{-0.5cm}
\caption{Optical spectral energy distribution (SED) of NGC\,5084 as detected by the 6dF survey and the APO/DIS observations. \emph{Top panel:} 4300-7000 \AA\ spectrum of the central 1.5 arcsec slit (APO/DIS, blue and red channels, in color) and the 6.7 arcsec radius fiber (6dF, black). \emph{Bottom left:} Detail of the H$\gamma$ spectral range (4200 -- 4550 \AA), showing the core as detected by 6dF and APO/DIS, as well as the North, South, East, and West subregions avoiding the core. See the labels on each spectra. \emph{Bottom right:} Same as previous for the 6400-6800 \AA (H$\alpha$) range. Vertical shadowed red lines represent the redshifted wavelengths of the typical absorption and emission lines in galaxies (H$\beta$, OIII, Mg, Na I, H$\alpha$), for reference.}
\label{fig:NGC5084_optical_spectra}
\end{center}
\end{figure*}

The optical spectral energy distribution of the core of NGC\,5084 was obtained by the 6dF Galaxy Survey \citep{jones+2004mnras355_747} in the 4000-7500 \AA\ range, using an aperture with a radius of $R=6.7$ arcsec around the core of the galaxy (see Fig.\,\ref{fig:NGC5084_optical_spectra}). The optical spectra reveal neither strong emission lines typical from central starbursts nor AGN associated broad-line emission (H$\beta$, [O\,III]), although H$\alpha$ and [N\,II] lines are detectable. Their optical line ratio is $\log_{10} \frac{[N\,II]\lambda6583}{[H\alpha]\lambda6563} = +0.51$. High values of this ratio are characteristic of high ionization power sources, (i.e., AGN, low mass evolved
stars), compared to active starburst galaxies where $[H\alpha] >> [N\,II]$ \citep{sanchezalmeida+2012apj756_163}.

\begin{figure*}[t!]
\begin{center}

\includegraphics[trim={0 0 70 0}, clip, height=8.5cm]{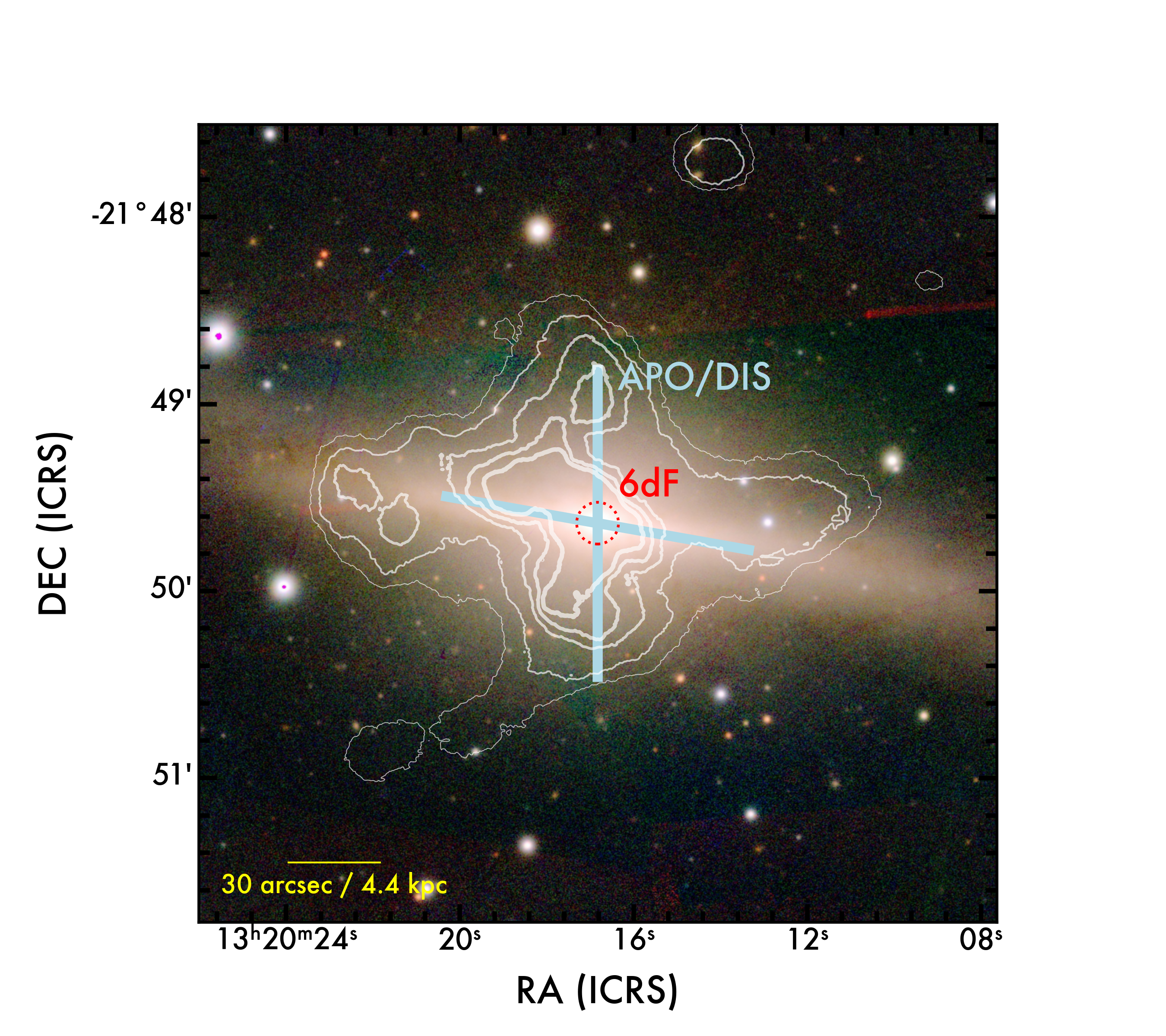}\includegraphics[trim={0 0 0 0}, clip, height=8.5cm]{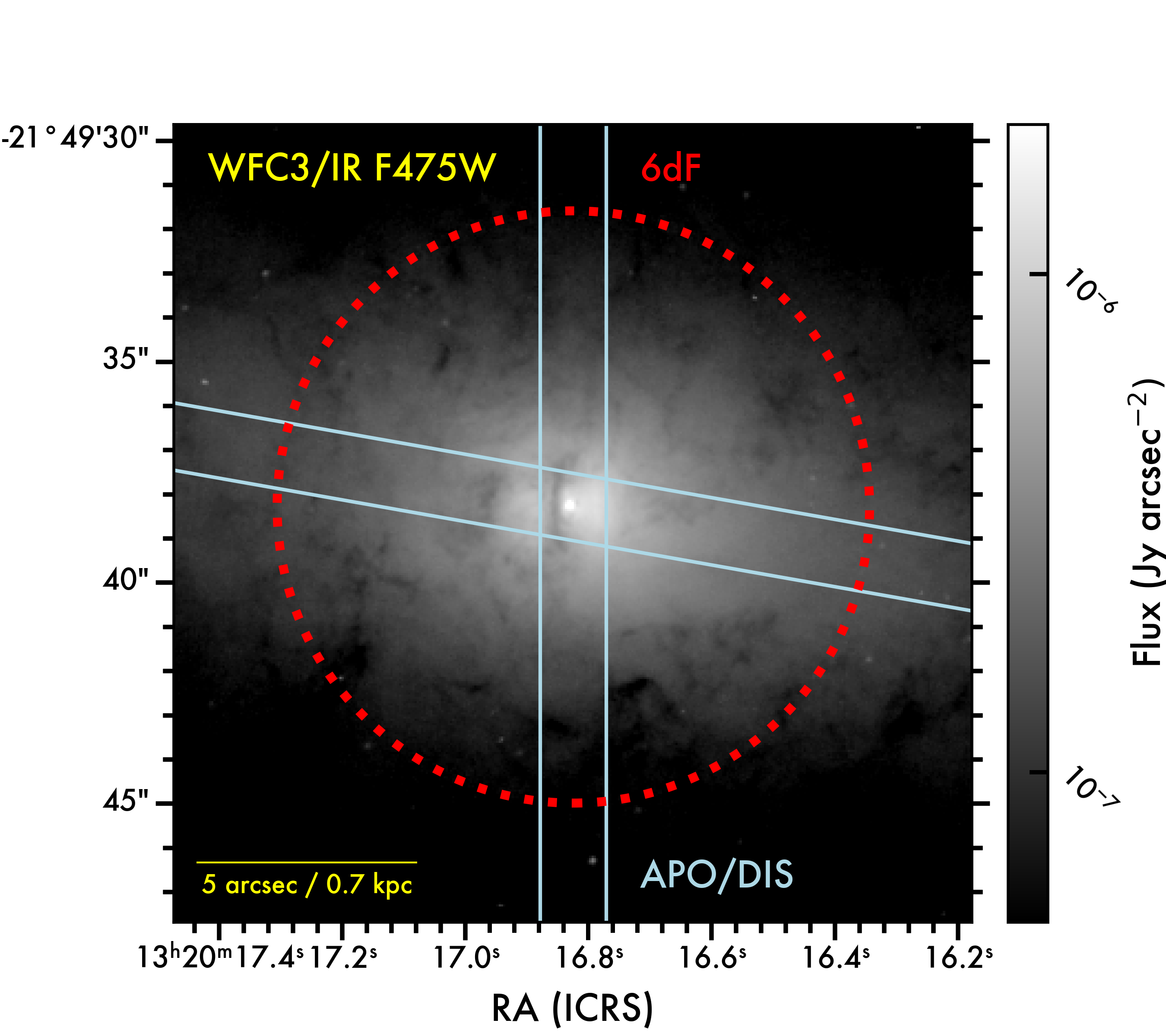}
\caption{Regions assigned for spectral analysis. \emph{Left panel:} White contours in the left panel represent the [2, 3, 5, 7, 10]$\sigma$ detection limits in the 0.3--2.0 keV band from Chandra/ACIS. The slit-shaped regions (1.5 arcsec wide) represent the APO/DIS optical spectra, defined to align with the major axis of the galaxy and the major axis of the circumnuclear disk. The circular core region, shown in the zoomed image, has a 3~arcsec radius. The RGB background image was generated using the $gri$ observations from Pan-STARRS. \emph{Right:} Close-up view of the core regions, showing the apertures for optical 6dF spectra ($R=3.4$ arcsec), and the APO/DIS slit-spectra. The background image represents the flux intensity in the F475W band from HST/WFPC2.}
\label{fig:NGC5084_spectra_regions}
\end{center}
\end{figure*}

Narrow emission lines from the AGN can erase the spectral absorption from the stellar component. In particular, we pay attention to Balmer absorption lines, the distinguishing feature of post-starburst population dominated galaxies \citep[also called E+A or K+A galaxies,][]{dressler+1983apj270_7}. In addition to the 6dF fiber-spectra, we analyzed the slit-spectra obtained with the APO/DIS
(see Sec.\,\ref{subsec:data_optical_spectra}
using five regions: (1) the central 1.5 arcsec from the core of the galaxy; (2-5) north, south, east, and west sides of the galaxy, resp. (not including the core). The results are shown in Fig.\,\ref{fig:NGC5084_optical_spectra}. The APO/DIS spectra do not reveal any signs of H$\alpha$ [$\lambda=6562.8\AA$], H$\beta$ [$\lambda=4861.35\AA$], or H$\gamma$ [$\lambda=4340.47\AA$] absorption, in any of the regions analyzed, suggesting that the stellar population at the core is not dominated by a classic post-starburst stellar population.

\section{Discussion} \label{sec:DIS}

\subsection{Main formation scenarios}
\label{subsec:discussion_main_formation_scenarios}
The present work has revealed a collection of remarkable features in the massive lenticular galaxy NGC\,5084, including:
\begin{enumerate}

    \item A hot gas halo with two components: a 17 kpc component perpendicular to the galactic disk, and a second parallel to it, detected in the $0.3-2.0$ keV band of \Chandra/ACIS (Secs.\,\ref{subsec:results_xray_ima}).

    \item A rotating ($V_{\rm rot} = 249.1^{+12.8}_{-9.6}$ km s$^{-1}$), edge-on ($i=71.2^{+1.8}_{-1.7}$$^{\circ}$), polar circumnuclear disk of $D=304^{+10}_{-11}$ pc in diameter, detected in absorption with \emph{Hubble} (WFPC2, WFC3) and confirmed through its CO(2-1) molecular emission in ALMA  observations (Secs.\,\ref{subsec:results_hst}, \,\ref{subsec:results_alma}).

    \item Two radio lobes located symmetrically at $\overline{R}=4.6\pm0.6$ kpc from the galactic core, aligned with the rotation axis of the circumnuclear disk, detectable in 6 and 20 cm (5 and 1.4 GHz) radio continuum emission observations with EVLA (Sec.\,\ref{subsec:results_radiopol}).


\end{enumerate}

\begin{figure*}[t!]
\begin{center}
\includegraphics[trim={0 0 0 0}, clip, width = \textwidth]{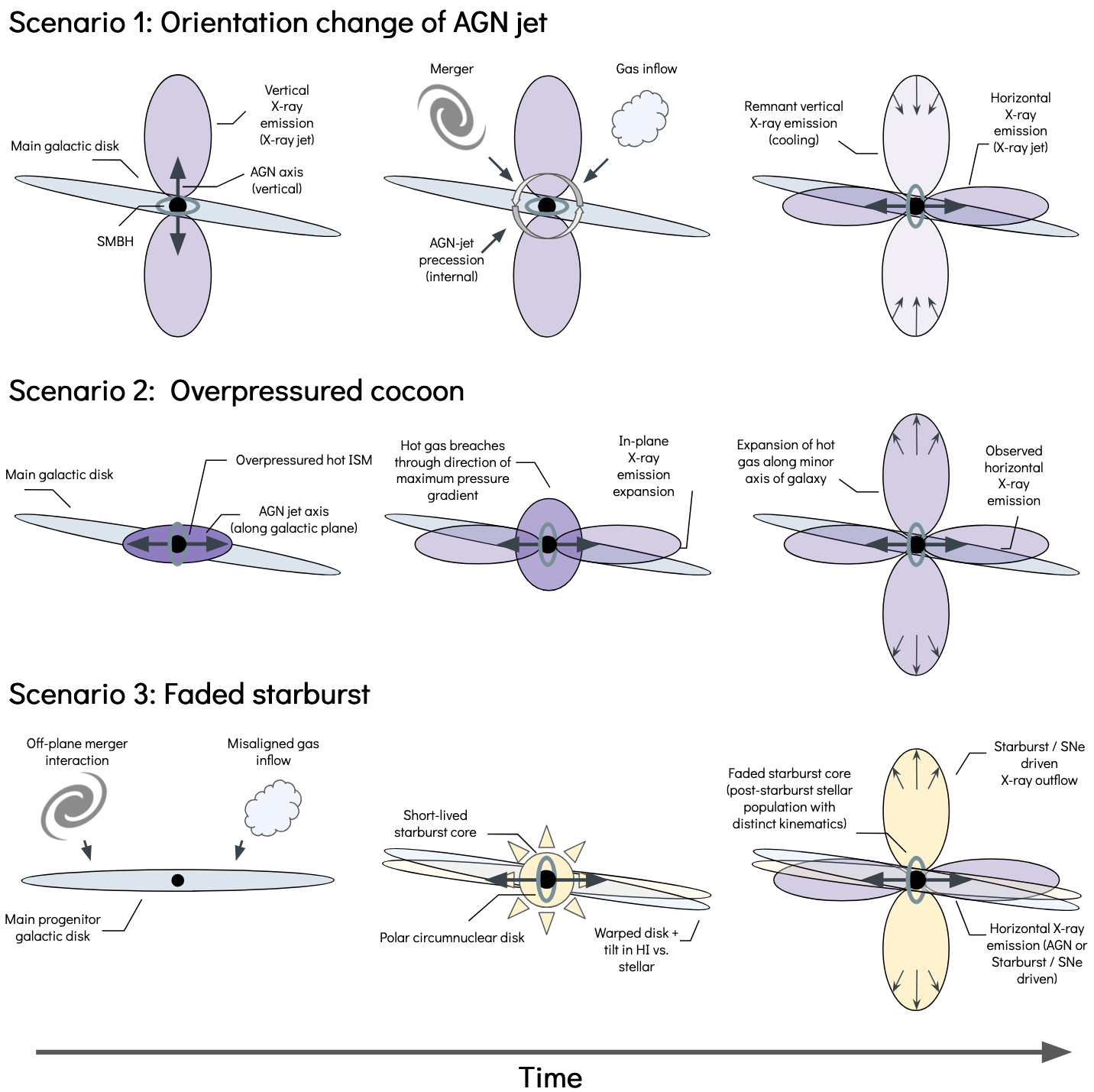}
\caption{Summary of the formation scenarios for the vertical X-ray emission of NGC\,5084. Cosmic time increases from left to right. \emph{Top row:} (1) Orientation change of AGN jet. Mergers, inflows, precession, and SMBH interactions can reorient the jet direction of an AGN over time; \emph{Central row:} (2) Overpressured cocoon. AGN-jet emission directed against the galactic disk can result in an expansion along the minor axis of the galaxy; \emph{Top:} (3) Faded starburst. Circumnuclear starbursts can generate galactic winds, expelling hot gas in the direction vertical to the galactic plane. However, this scenario requires active star formation or at least a relatively young population of stars at the core, which is not observed (see Sec.\,\ref{subsec:discussion_fadedstarburst}).}
\label{fig:NGC5084_scenarios}
\end{center}
\end{figure*}

The presence of (1) a rotating, circumnuclear disk and (2) two radio lobes aligned with the axis of the circumnuclear disk is evidence for a SMBH at the core of NGC\,5084, and adds support for extant AGN activity reported by previous authors \citep{kewley+2001apj556_121}. Following the prescriptions of \citet{smith+2021mnras500_1933}, we estimated an SMBH mass of $\log_{10}\left( \frac{M_{\rm BH}}{M_{\odot}} \right) = 7.66^{+0.21}_{-0.15}$. The measured SMBH mass is compatible with the SMBH mass distribution in the local Universe \citep{vika+2009mnras400_1451, smith+2021mnras500_1933}. Remarkably, all evidence suggests that the AGN jet axis is pointing towards the galactic disk plane, instead of being perpendicular to the main galactic disk. Based on the multi-wavelength morphological information, we propose three main formation scenarios for NGC\,5084, summarized in Fig.\,\ref{fig:NGC5084_scenarios}:

\begin{enumerate}
    \item \emph{Orientation change of AGN jet:} In this scenario, the vertical X-ray emission was generated in a previous orientation of the AGN jet, which changed dramatically due to external (merger, gas inflow) or internal factors \citep[AGN precession, ][]{britzen+2018mnras478_3199, britzen+2023apj951_106}.


    \item \emph{Overpressured coccoon:} NGC\,5084's AGN jet pointing towards the disk heats up the ISM in the core of the galaxy. The hot gas escapes through the direction of maximum pressure gradient, that is, the minor axis of the disk, generating an expansion visible as the vertical component of the X-ray emission \citep[see Fig.\,5 in][]{capetti+2002aap394_39}.

    \item \emph{Faded starburst:} NGC\,5084 evolved through one or more gas-rich accretion events, driving gas into the core, which in turn: 1) triggered a starburst-driven galactic wind \citep{hodgeskluck+2020apj903_35} and 2) activated the AGN and its central SMBH, generating the observed horizontal X-ray emission.
\end{enumerate}

These scenarios are idealized; in reality, a mixture of these cases may be responsible for the observed phenomena. In particular, the HI tilt and the disk warp of NGC\,5084 suggest that the galaxy suffered a major merger at some point in its past. Rather than directly eliminating other scenarios, mergers serve as a potential triggering factor in two of the other cases (orientation change and faded starburst). The observational evidence for merger activity is gathered in Sec.\,\ref{subsec:discussion_merger}. Secs.\,\ref{subsec:discussion_fadedstarburst}, \ref{subsec:discussion_cocoon}, and \ref{subsec:discussion_realignment} discuss each of the scenarios presented in Fig.\,\ref{fig:NGC5084_scenarios}.

\subsection{Signs of merger activity}
\label{subsec:discussion_merger}

The size ($D=304^{+10}_{-11}$ pc) and rotation speed ($V_{\rm rot} = 249.1^{+12.8}_{-9.6}$ km s$^{-1}$) of the circumnuclear disk detected with \emph{Hubble} and ALMA observations is compatible with the NLR dust disks that surround the AGN torus \citep[][]{ramosalmeida+2017nat1_679}. However, its perpendicular orientation to the rest of the galaxy poses an important question regarding its kinematic origin. \par

The presence of polar nuclear rings is a tracer of past accretion events in disk galaxies \citep{anantharamaiah+1996apj466_13}. Observations of these structures in NGC\,2685 and IC\,1689 revealed a stellar population different from that of the rest of the galaxy, suggesting the formation by an accretion event followed by a starburst \citep{silchenko1998aap330_412}. Surveys of galactic nuclear rings \citep{comeron+2010mnras402_2462} have also found that the distribution of position angles of nuclear rings shows no correlation with that of the rest of their host galaxies, supporting an accretion scenario for their formation. These findings indicate that the orientation of the inflow gas does not depend generally on the orientation of the host galaxy, but rather on the distribution of external material to be accreted.

Alternative mechanisms have been proposed to generate misaligned or even polar nuclear disks via galactic outflows without the need for gravitational interactions. Simulations have proven that star formation, supernovae, and AGN feedback can eject gas above the galactic plane \citep{emsellem+2015mnras446_2468, renaud+2015mnras454_3299}. Eventually this gas falls back to the core with a certain angular momentum, creating polar gas rings or disks \citep{combes2017inproceedings_Di}. In this scenario, the X-ray emission perpendicular to the disk observed in NGC\,5084 would be hot gas precipitating back into the core of the galaxy while cooling down, and then eventually transforming into the observed circumnuclear dust disk. However, despite NGC\,5084's extensive vertical X-ray emission, its HI emission does not show signs of filamentary structures expected with infall to the core of the galaxy  \citep{zheng+2022afz22_085004}, raising questions about this possibility.

\begin{figure*}[t!]
\begin{center}
\includegraphics[trim={0 0 0 0}, clip, width=\textwidth]{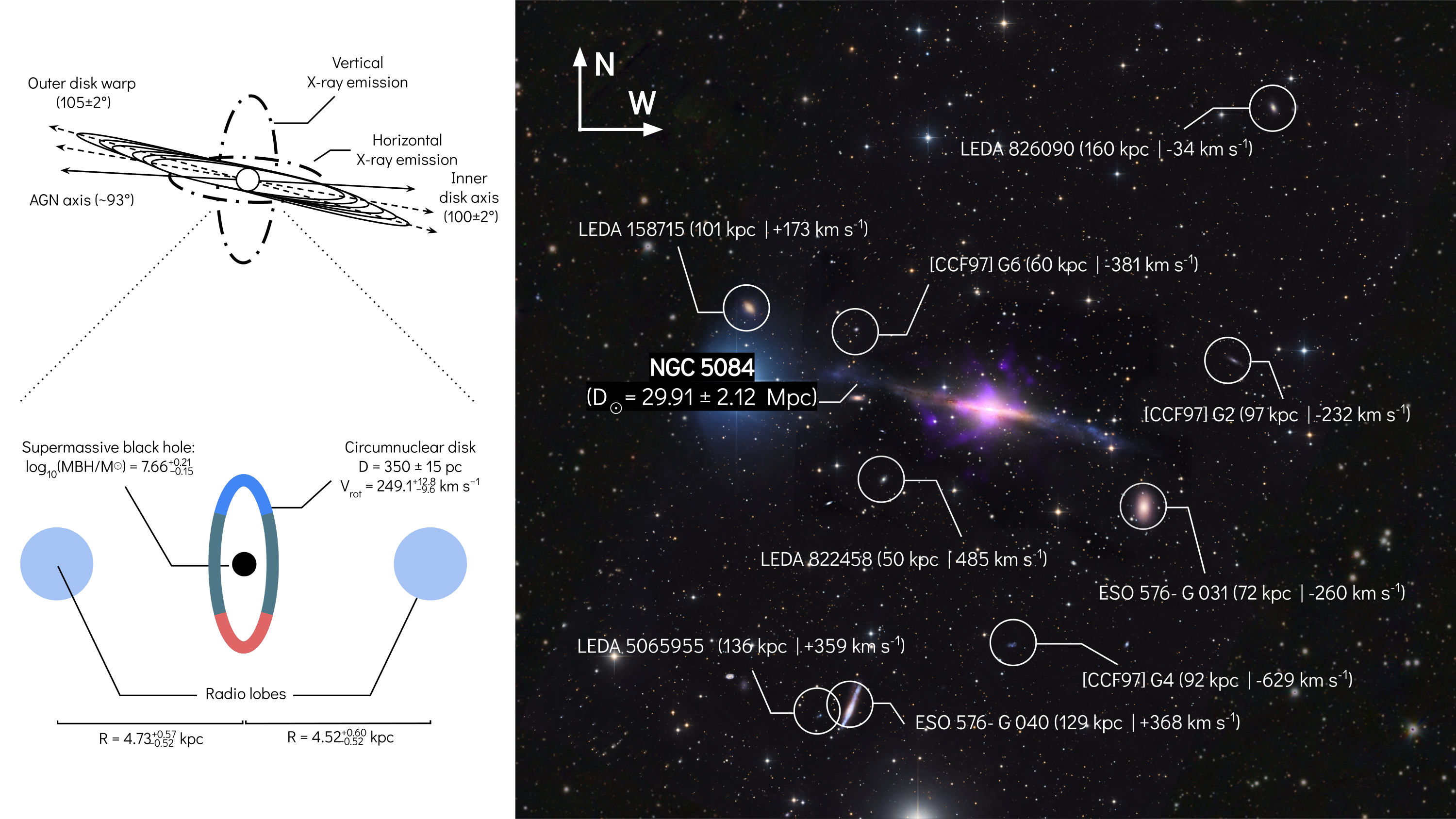}
\caption{NGC\,5084 morphological and environment scheme. \emph{Top left:} Main components at scales from 5 to 20 kpc, including the vertical and horizontal X-ray emission, its inner stellar disk, and the warped outer disk. \emph{Bottom left:} NGC\,5084 components at 0.1 to 5 kpc scales, including the symmetric radio lobes, the polar circumnuclear disk, and the supermassive black hole at its center. \emph{Right panel:} Environment of NGC\,5084. Each circle represents a satellite galaxy around NGC\,5084. Original data from \citet{carignan+1997aj113_1585}.}
\label{fig:NGC5084_environment}
\end{center}
\end{figure*}

\citet{gottesman+1986mnras219_759} interpreted that the observed alignment between the inner regions of NGC\,5084's stellar disk (or lens) and the neutral atomic hydrogen (HI) gas disk ruled out an accretion scenario. However, later works \citep{elichemoral+2018aap617_113} have demonstrated that lenses can result after major mergers in lenticular galaxies, implying that their presence is not sufficient to rule out mergers. While the presence of a polar circumnuclear disk associated with an AGN can be explained by internal evolution mechanisms \citep[outflow gas cooling and subsequent precipitation onto the core,][]{combes2017inproceedings_Di}, multiple forms of evidence point to some type of accretion event (gas inflow, minor, or major merger) in the recent past of NGC\,5084, including:

\begin{enumerate}
\item The observed warp in the outer disk ($\Delta\Theta\sim5^{\circ}$) \citep{zeilinger+1990mnras246_324}.

\item The misalignment of the HI gas disk with respect to the stellar disk \citep{gottesman+1986mnras219_759,zheng+2022afz22_085004}.
\item The detection of an antitruncated surface brightness profile \citep{comeron+2012apj759_98}, an excess of light in the outskirts potentially generated by past merger events \citep{younger+2007apj670_269, borlaff+2014aap570_103}.

\item The presence of nine satellite galaxies in its environment, with an apparent predominance of satellites in retrograde orbits, more likely to survive accretion \citep[see Fig.\,\ref{fig:NGC5084_environment}]{carignan+1997aj113_1585}.

\end{enumerate}

Based on this observational evidence in the cosmological context of massive S0 galaxies in the local Universe, we conclude in Sec.\,\ref{subsec:origin_of_NGC5084} that the most likely scenario is that NGC\,5084 sustained a major merger and/or multiple minor mergers since $z\sim1$.

\subsection{Faded starburst scenario}
\label{subsec:discussion_fadedstarburst}

Galactic superwinds \citep[][]{heckman+1993inproceedings_455} are generated by  radiative and kinetic luminosity provided by supernovae and winds from massive stars in starburst regions. The collisions of the ejecta (shocks) heat up the surrounding gas to $T\gtrsim10^{8}$ K, filling the ISM and preferentially expanding towards regions in which the pressure gradient is more steep (that is, the perpendicular direction to the galactic disk), generating a "bubble" of hot X-ray emitting gas. Thus, galactic outflow tend to be perpendicularly aligned to the disk, explaining the bipolar component observed in X-ray.

Observationally, $\sim$50\% of star-forming galaxies show signs of winds \citep[][]{rubin+2014apj794_156}. Extreme examples would be M\,82 \citep{shopbell+1998apj493_129,heckman+2017incollection_2431} and the giant X-ray wind cone of NGC\,3079 \citep[$R\sim3$ arcmin, 16.3~kpc,][]{hodgeskluck+2020apj903_35}. In contrast to typical star-forming galaxies, NGC\,5084 is a lenticular galaxy with a low SFR \citep[$0.128\pm0.016$ M$_{\odot}$ yr$^{-1}$,][]{osullivan+2018aap618_126}, showing no evident emission lines characteristic of active star formation \citep[][see Fig.\,\ref{fig:NGC5084_optical_spectra}]{moustakas+2006apj164_81}. If the observed X-ray outflow of hot gas observed in NGC\,5084 \Chandra/ACIS is a galactic superwind, the star formation burst that generated it must have been a short event, and must have been completely quenched in the present. Moreover, X-ray outflows from galaxies that have already stopped their starburst activity tend to extend for periods of time no longer than a few tens of Myr \citep{mcquinn+2018mnras477_3164}.  In this scenario, the stellar population in the core should be relatively recent (post-starburst).

\begin{figure*}[t!]
\begin{center}
\includegraphics[trim={0 20 40 0}, clip,  height = 7.5cm]{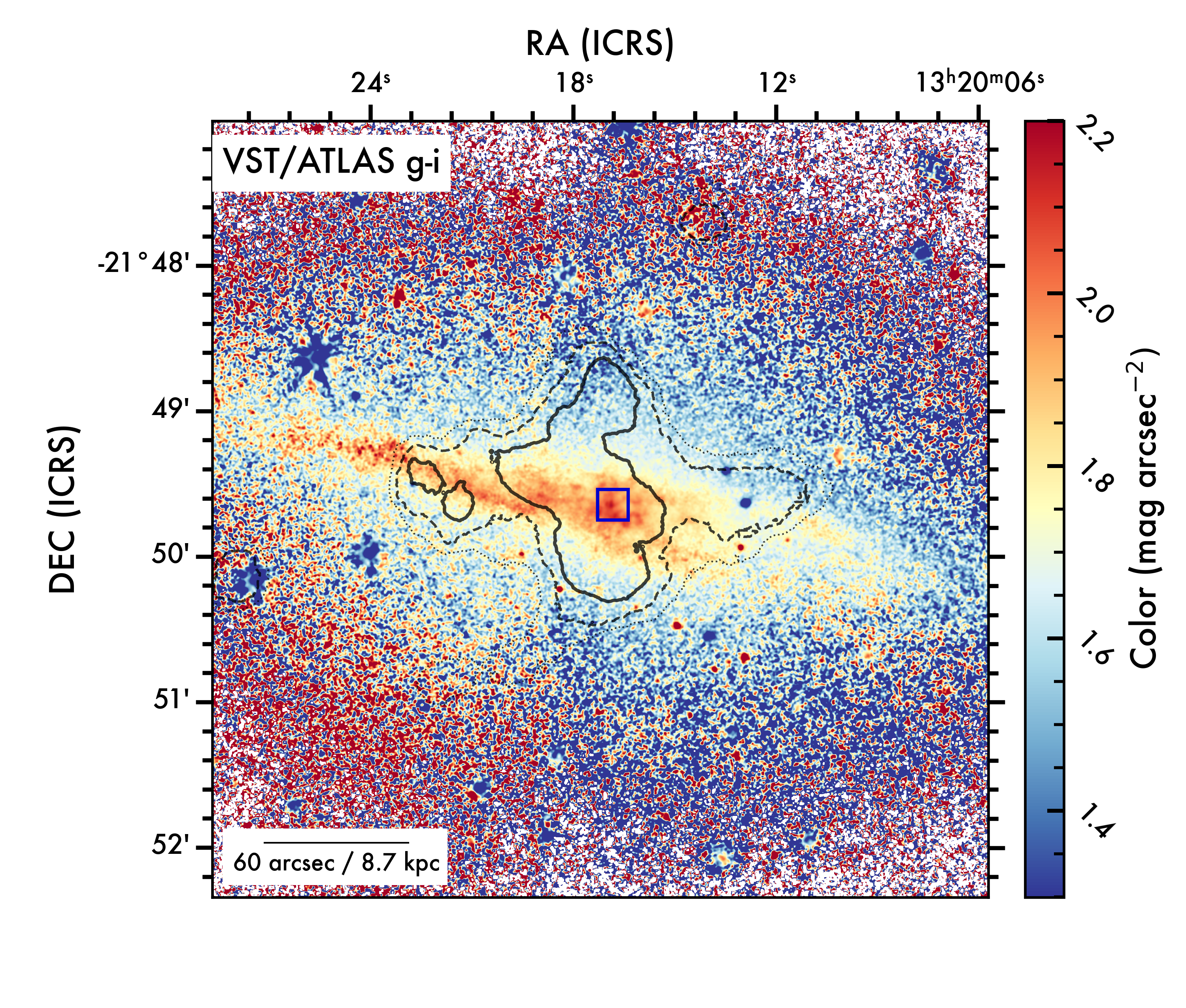}
\includegraphics[trim={30 20 0 0}, clip,  height = 7.5cm]{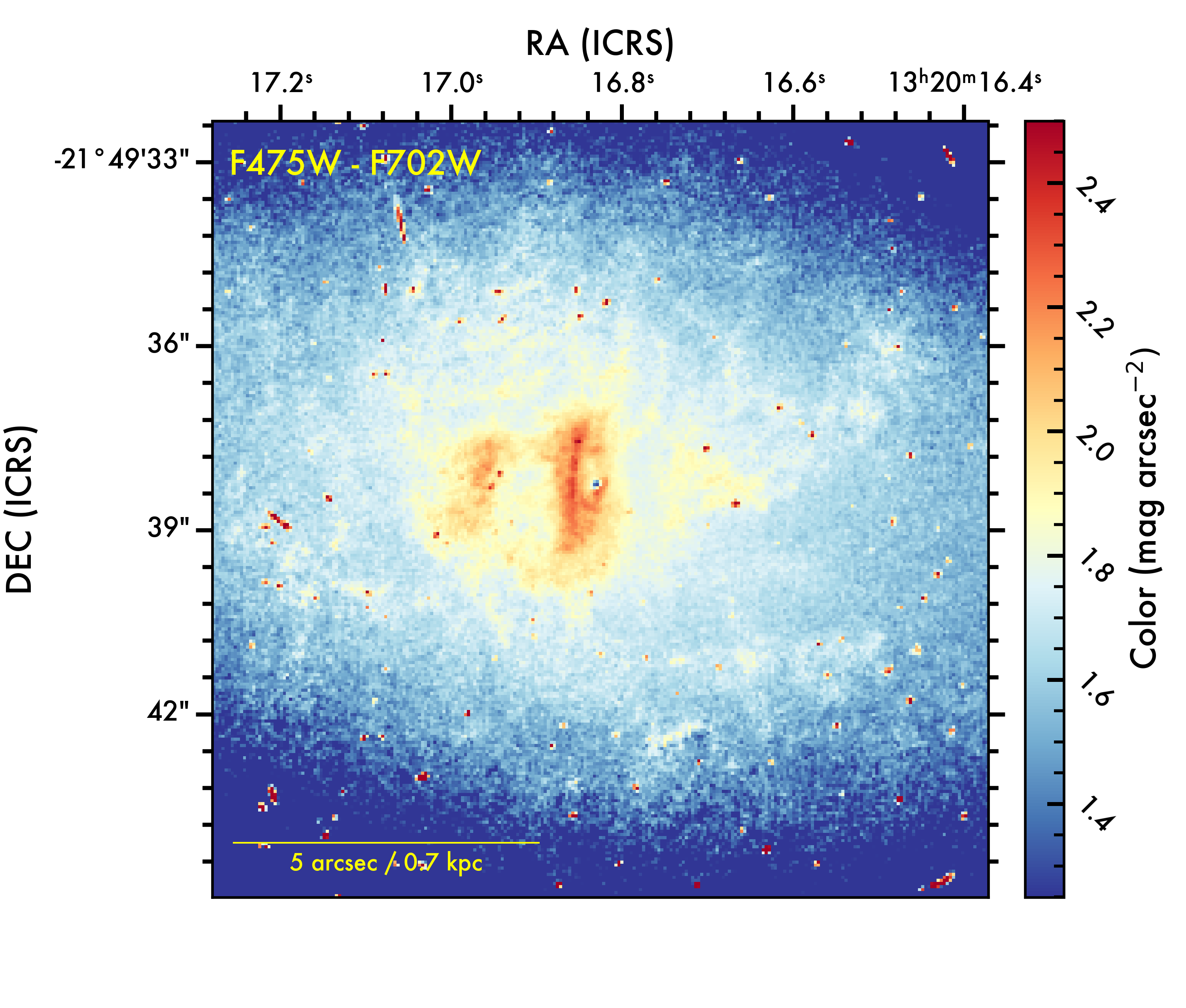}
\caption{NGC\,5084 color maps. \emph{Left panel:} Ground-based VLT/ATLAS $g-i$ maps. Black contours represent the X-ray emission (see Fig.\,\ref{fig:NGC5084}). Dark blue square in the core represents the FOV in the right panel. \emph{Right panel:} F475W - F702W color from \emph{Hubble} WFPC2/WFC3 photometry. The circumnuclear disk is clearly visible in red. See the colorbar for reference.}
\label{fig:NGC5084_color}
\end{center}
\end{figure*}

However, this hypothesis is not supported by the spectro-photometric information available. First, the optical color diagrams of the central regions of NGC\,5084 (based on $g-r$ VST/ATLAS images\footnote{VST/ATLAS: \url{http://osa.roe.ac.uk/}}, see Fig.\,\ref{fig:NGC5084_color}) do not reveal a bluer emission at lower galactocentric radius but rather indicate a non-uniform internal dust absorption throughout the disk. Dust reddening is expected in the core, but even HST observations reveal that the general color of the diffuse emission between dust filaments is relatively red ($g-i\gtrsim1.4$).
Secondly, the optical spectra (see Sec.\,\ref{subsec:results_Optical_spectra}, Fig.\,\ref{fig:NGC5084_optical_spectra}) do not present any Balmer absorption lines typical of post-starburst objects \citep[also called E+A or K+A galaxies,][]{dressler+1983apj270_7}, not even in regions away from the core, where the AGN could dominate the emission. Given that the post-starburst phase lasts about 300\,Myr, the lack of a Balmer absorption line signal in the optical spectra suggest that the age of the stellar population in the core is older than a Gyr.

\subsection{Overpressured cocoon}
\label{subsec:discussion_cocoon}

Scenario 2 in Fig.\,\ref{fig:NGC5084_scenarios} describes an adaptation of the mechanism proposed by \citet{capetti+2002aap394_39} to generate X-shape radio sources: a bubble of hot expanding gas inside an elliptical distribution of denser gas will expand along the direction of maximum pressure gradient, that is, the minor axis of the galaxy  \citep{blandford+1974mnras169_395}. The surrounding galactic ISM collimates the expanding vertical exhaust, generating well-defined wings. Notably, this type of object has been predominantly detected in radio wavelengths \citep[X-shaped radio galaxies,][]{hodgeskluck2011thesis, giri+2024sci11_1371101}.  X-ray observations of such objects tend to reveal diffuse and elliptical hot gas halos \citep{hodgeskluck+2010apj710_1205}, cavities \citep{hodgeskluck+2010apj717_37}, or hard X-ray knots around the radio emission \citep{hodgeskluck+2012apj746_167}. The cross-shaped X-ray morphology observed in NGC\,5084 is unique in this sense.

Unfortunately, the systematic effects observed in radio observations along the vertical direction (Sec.\,\ref{subsec:results_radiopol}) prevent the identification of additional radio lobes along the minor axis of the galaxy, challenging the ability to determine whether NGC\,5084 also has an X-shape radio morphology. Given the detected radio-lobes, NGC\,5084 would be classified between high-luminosity LINER/Seyferts and FR~0 galaxies \citep[see Table \ref{tab:Radio_Lobes},][]{baldi2023aap31_3}, emitting $\sim$4--5 orders of magnitude less than the X-shaped radio galaxies from \citet{cheung+2009apj181_548}. This classification correlates with the prediction from \citet{hodgeskluck2011thesis} that X-shaped sources should be decaying AGN jets.

\subsection{AGN re-alignment}
\label{subsec:discussion_realignment}

AGN jets can be almost randomly oriented with respect to their circumnuclear disks \citep{schmitt+2002apj575_150}  and it is expected that severe warps can be present between the orientation of the NLR and the BLR \citep{lawrence+2010apj714_561}, and by extension, with the main galactic disk \citep{clarke+1998apj495_189, nagar+1999apj516_97, kinney+2000apj537_152}. Potential causes for this misalignment include (1) a time-variation in the orientation of the jet-generator; or (2) the presence of pressure gradients in the trajectory of the jet \citep{gallimore+2006aj132_546}. The remarkable symmetry of the 6 cm continuum radio lobes detected with EVLA (see Fig.\,\ref{fig:NGC5084_EVLA}) compared to the core of the galaxy confirms that the AGN jet is currently aligned with the axis of the circumnuclear disk and pointing towards the galactic plane. However, this configuration does not need to be constant throughout the whole history of NGC\,5084.

In addition to the overpressure cocoon scenario, \citet{giri+2024sci11_1371101} reviews another mechanism to generate X-shaped sources: the reorientation of the jet. Observational works have revealed that AGN jets present signs of restarted activity in different directions  \citep{saripalli+2013mnras436_690, nandi+2021apj908_178}. In fact, jet reorientation by the action of binary supermassive black holes is one of the proposed mechanisms to explain the variable activity of blazars \citep{britzen+2018mnras478_3199}. The presence of a binary SMBH can be detected by searching for time variations on the radio emission at pc-scale of AGN jets \citep{jiang+2023apj959_11}. Since secondary SMBHs are likely to be accreted in merger events, their study may be critical in establishing observational constraints on galaxy evolution \citep{yu2002mnras331_935}. Binary SMBHs that are separated at sub-pc scales can generate changes in the jet orientation over periods of several Myr \citep{giri+2024sci11_1371101}, sufficiently short so that the fossil X-ray emission from the previous orientation is still visible \citep[$\sim10-20$ Myr]{zubovas+2022mnras515_1705}.

\subsection{NGC\,5084 in a cosmological context}
\label{subsec:origin_of_NGC5084}

Observational works suggest that at least 75\% of AGN host galaxies have recently interacted or are on-going mergers \citep{keel+2012mnras420_878}. This fraction contrasts with $\Lambda$CDM simulation-based works, which predict that only half ($\sim$55\%) of the luminous AGNs ($L_{\rm bol} > 10^{45}$ erg s$^{-1}$) are in some type of gravitational interaction, suggesting that mergers are an important factor but not the only one to trigger AGNs \citep{byrnemamahit+2024arXiv2402.05196}.

In gravitational interactions, the galactic gas loses angular momentum and precipitates into the core regions of the galaxy \citep{hopkins+2009apj691_1168}. This gas is often consumed in starbursts but also reaches the SMBH, which fuels AGN activity \citep{dimatteo+2005nat433_604, springel+2005mnras361_776}. Thus AGN activity and outflows can be triggered by gravitational interactions. However, AGNs are active in periods of $0.2-2\times10^{5}$ yr \citep{keel+2012mnras420_878}, a timescale much shorter than typical starbursts \citep[$\tau=10^{7}$ yr,][]{heckman+1993inproceedings_455}. These timescales imply that simultaneous observations of an AGN and the starburst may not be common, despite being triggered by the same event.

AGNs are one of the multiple mechanisms to shut down star formation in galaxies, but not the only one. From a cosmological perspective, the transformation from spirals to S0s had to take place either through internal processes or through external interactions with other galaxies. In the internal scenario, spiral galaxies would have shut down their star formation, either through gas stripping in the disc \citep{laurikainen+2010mnras405_1089}, removal of their gas \citep[starvation or strangulation,][]{larson+1980apj237_692}, harassment \citep{moore+1996nat379_613}, strong and sustained star formation \citep{kormendy+2004araa42_603} or AGN feedback \citep{chen+2020apj897_102}. \par

Contrary to the classical galaxy evolution scenario, many studies propose that galactic mergers can produce well-defined disk remnants \citep{hopkins+2013mnras430_1901,elichemoral+2018aap617_113} or even spiral galaxies \citep{athanassoula+2016apj821_90,peschken+2017mnras468_994}. In fact, bulge/disk decompositions suggest that the low-luminosity S0s were formed by secular processes, while the bright (more massive) S0s may have been formed through major mergers \citep{barway+2009mnras394_1991,frasermckelvie+2018mnras481_5580}. This second class includes supermassive lenticular galaxies like NGC\,5084.\par

Another example of such phenomena is the lenticular galaxy NGC\,5252. Classified as a Seyfert 1.9 galaxy \citep{argyle+1990mnras243_504, osterbrock+1993apj414_552}, NGC\,5252 presents a large-scale ionization bi-cone \citep{tadhunter+1989nat341_422} detectable in [OIII] emission line at $\lambda=5007\AA$ that extends for $R\sim20$ kpc. \emph{Hubble} Space Telescope and Fabry--P\'{e}rot spectrograph observations of NGC\,5252 by \citet{morse+1998apj505_159} revealed different kinematic components, including an inclined circumnuclear gas disk with a diameter of 3 kpc, suggesting that NGC\,5252 underwent a galaxy merger in its past history. This scenario is supported by observational evidence of the presence of both a supermassive (main) and an intermediate (accreting) mass black hole in NGC\,5252, both active and emitting in radio \citep{kim+2015apj814_8, kim+2017apj844_21, yang+2017mnras464_70}. Moreover, recent results presented by \citet{wang+2024arXiv2401.09172} show a strong ($\sim20^{\circ}$) misalignment between the X-ray emission in soft bands (0.3--2.0 keV) and the optical major axis of the galaxy. The case of NGC\,5252 shows that the combination of observational evidence such as the presence of off-axis morphological components and AGN-related features is one strategy to shed light on the formation pathways of specific galaxies.

Processes related to the environment (i.e., gas evaporation) may only play a significant role in the formation of intermediate-to-low mass S0 galaxies located in clusters \citep{vollmer+2012aap537_143, cerulo+2017mnras472_254}. However, half of the S0 galaxies reside in groups or on the field, with very few cases of isolated S0s \citep{khim+2015apj220_3}. NGC\,5084 is the main galaxy of its own group, associated with the Virgo II southern extension of the Virgo cluster \citep{tully1982apj257_389}, and it is one of the most massive lenticular galaxies in the Local Universe \citep[more massive than 98\% of the S0 galaxies]{ohlson+2024aj167_31}. Taking into account the multi-wavelength morphological and spectroscopic evidence analyzed in this work, \emph{NGC\,5084 is perhaps one of the most clear cases of an S0 galaxy assembled through hierarchical accretion.}

\section{Conclusions} \label{sec:CON}

No single observatory has the capability to unveil the full picture of galaxy evolution. Multi-messenger studies\footnote{The term \emph{multi-messenger}, as opposed to \emph{multi-wavelength} includes new tracers of astrophysical phenomena, such as gravitational waves, not associated to the detection of light.} are the key to detecting the different layers of the galactic structure and inferring their formation mechanisms, and in understanding how the Universe transformed from relatively simple structures in the early Universe to the current observed complexity across Cosmic Time  \citep{beckman2021book}. \par

In this work, we have explored a single object, NGC\,5084, a supermassive lenticular galaxy combining observations obtained with Chandra X-ray observatory (X-ray), Hubble Space Telescope, EVLA, and ALMA, with additional support from 6dF, APO, and VST/ATLAS observations. NGC\,5084 displays a extended cross-pattern X-ray feature spanning 17 kpc in the vertical direction, and 22 kpc in the galactic plane, enclosing a $D=304^{+10}_{-11}$ pc polar circumnuclear disk at its center. At $\overline{R}=4.6\pm0.6$ kpc from the core in the galactic plane direction, two symmetric radio lobes reveal the effects of an AGN in the core of NGC\,5084, whose axis points towards the disk and is aligned with the rotation axis of the circumnuclear disk. The detection of the CO(2-1) molecular emission line in the circumnuclear disk allows a measurement of its rotation velocity (V$_{\rm rot} = 249.1^{+12.8}_{-9.6}$ km s$^{-1}$) and the mass of the SMBH at its core ($\log_{10}\left( \frac{M_{\rm BH}}{M_{\odot}} \right) = 7.66^{+0.21}_{-0.15}$). \par

The detection of elongated X-ray lobes is insufficient to single-handedly discriminate between AGN- and starburst-powered jet outflow as possible origins for the structure, or to unequivocally exclude other scenarios. Supplemental data that can be used to search for kinematic signatures of outflows and the presence of ionization cones are essential to favoring or rejecting models.
The horizontal (in-plane) component of the X-ray emission is aligned with both the rotation axis of the newly discovered circumnuclear disk and the symmetric radio lobes, identified by \citet{irwin+2019aj158_21} as part of a radio AGN jet. Taking into account all available observational evidence consolidated in this paper (see Fig.\,\ref{fig:NGC5084_scenarios}), we consider three potential hypotheses for the formation of the observed cross-shaped X-ray emission: (1) it is the remnant of a re-oriented AGN; (2) it is an outflow generated by an overpressured cocoon of hot gas powered by the AGN pointing into the dense ISM within midplane of the galactic disk; or (3) it is part of a faded starburst at the core of the galaxy. Spectroscopic observations on the core of NGC\,5084 do not support the latter scenario, given the lack of spectral evidence for recent (massive stars) or on-going star formation in the core.  Combining the new observational evidence presented in this paper with previous analysis based on environment and morphology (see Sec.\,\ref{sec:DIS}), we conclude that NGC\,5084 is with high probability the remnant of at least one merger in the past, which is actively accreting its multiple satellites.

The discovery of NGC\,5084 circumnuclear disk and hot gas cross-shaped emission was based entirely on archival observations, some obtained almost three decades ago (see Table \ref{tab:Observations}), spanning seven orders of magnitude in wavelength range, and covering physical scales from several pc into the core to tens of kpc into the halo. This remarkable combination of observations reflects the importance of one of the foundational activities that pave the way to transformative science as recognized by the Decadal Survey on Astronomy and Astrophysics 2020 \citep{2021pdaa.book.....N}: the support of data archives. Curated archives are the pillar to future discoveries based on advanced techniques -- some of them tested in the present work, but most of them yet to be developed -- which will enable the detection of previously unseen phenomena from both old and new observations.

\begin{acknowledgments}
The authors thank the anonymous referee for the provided input that helped to improve this publication significantly. The list of Chandra datasets, obtained by the Chandra X-ray Observatory, are contained in~\dataset[DOI: 10.25574/cdc.273]{https://doi.org/10.25574/cdc.273}. The HST data presented in this paper were obtained from the Mikulski Archive for Space Telescopes (MAST) at the Space Telescope Science Institute. The specific observations analyzed can be accessed via \dataset[DOI: 10.17909/j04e-sq50]{https://doi.org/10.17909/j04e-sq50}. A.B. acknowledges the tireless support from
%
the CXC Helpdesk team. Without your dedication, this project would have not been possible. Thanks to Dr. Carlos G\'{o}mez Guijarro for this deep review of this manuscript and insightful advice. This research has made use of data obtained from the Chandra Data Archive and the Chandra Source Catalog, and software provided by the Chandra X-ray Center (CXC) in the application packages \texttt{CIAO} \citep{fruscione+2006inproceedings_62701V} and \texttt{Sherpa} \citep{freeman+2001inproceedings_76}.
Support for this work was provided by the National Aeronautics and Space Administration through Chandra Award \#24610329 issued by the Chandra X-ray Center, which is operated by the Smithsonian Astrophysical Observatory for and on behalf of the National Aeronautics Space Administration under contract NAS8-03060.
This work was co-authored by an employee of Caltech/IPAC under Contract No. 80GSFC21R0032 with the National Aeronautics and Space Administration. N. C. and S. D. C.'s research are supported by an appointment to the NASA Postdoctoral Program at the NASA Ames Research Center, administered by Oak Ridge Associated Universities under contract with NASA.
\end{acknowledgments}

%

\vspace{5mm}
\facilities{HST (WFPC2, WFC3), \Chandra\ (ACIS), ALMA, EVLA}


\software{\texttt{aplpy} \citep{2012ascl.soft08017R}, \texttt{astropy} \citep{collaboration+2018aj156_123, collaboration+2013aap558_33, collaboration+2022apj935_167}, \ciao, \texttt{LIRA} \citep{donath+2022inproceedings_98}\footnote{\texttt{pyLIRA:} \url{https://github.com/astrostat/pylira}}, \texttt{Matplotlib} \citep{hunter2007sci9_90}, \texttt{VorBin} \citep{cappellari+2003mnras342_345}, \texttt{SAODS9} \citep{2000ascl.soft03002S}}


\appendix

\section{Extended X-ray Emission Detection Tests}
\label{Appendix:xray_tests}

In this Appendix we expand on the technical aspects of \SAUNAS's \citep{borlaff+2024apj967_169} adaptively smoothed surface brightness maps (Sec.\,\ref{Appendix:xray_SNR}) for NGC\,5084. The quality checks described below provide additional verification of the detection of  extended X-ray emission in NGC\,5084, independent of both PSF deconvolution (Sec.\,\ref{Appendix:Xray_noPSFdeco_test}) and Voronoi binning (Sec.\,\ref{Appendix:Xray_novoro_test}).

\subsection{Signal Detection in Extended Sources}
\label{Appendix:xray_SNR}

One of the main advantages of \SAUNAS's spatial binning application to the X-ray event maps is the improvement in detectability of extended low surface brightness sources. Extended emission sensitivity limits in binned maps are considerably fainter than the reference point source sensitivity of \Chandra/ACIS\footnote{\Chandra/ACIS proposer guide: \url{https://cxc.harvard.edu/proposer/POG/html/chap6.html}} ($4\times10^{-15}$ erg cm$^{-2}$ s$^{-1}$ in 10$^4$ s, at 0.4--6.0 keV). In general, the sensitivity limit of \Chandra/ACIS ($f_{\rm lim}$) at a given signal-to-noise ratio ($\sigma$) detection level is a function of photon energy ($E$), binning area ($A$), exposure time ($t$), and equivalent effective area \citep[$M$, which has a complex variation with energy, see Fig.\,6 in][]{evans+2010apj189_37}:

\begin{equation}
\label{eq:F_lim_Xray}
f_{\rm{lim}}\,[\rm{erg}\,\rm{cm}^{-2}\,\rm{s}^{-1}\,\rm{arcsec}^{-2} ] =\\ 1.6\cdot10^{-9}\,[\rm{erg} \, \rm{keV}^{-1}] \times\frac{\emph{E\,[\rm{keV}]}\,\sigma^2}{\emph{A}\,[arcsec^2]\,\emph{M}\,[cm^2]\,\emph{t}\,[s]}
\end{equation}

At $E=1$ keV and $t=10$~ks, the equivalent exposure\footnote{\ciao/\Chandra\ exposure maps: \url{https://cxc.cfa.harvard.edu/ciao/threads/expmap_acis_single/}} ($\emph{M}\,\emph{t}$) is approximately $3.8\times10^{6}$ cm$^{2}$ s$^{1}$. For a point source detection of $\sigma=3$, with flux integration over an area associated with the PSF at the center of the ACIS detector ($FWHM\sim1.1"$), Eqn.\,\ref{eq:F_lim_Xray} implies a limiting sensitivity of $f_{\rm{lim}} = 4\times10^{-15}$ erg cm$^{-2}$ s$^{-1}$, equal to the reference instrument point source sensitivity limits.

For extended sources, increased aperture size results in fainter limiting sensitivities. For example, binning on areas of $25"\times25"$ produces a $\sigma=3$ detection of an extended source with average surface brightness of $5\times10^{-18}$ erg cm$^{-2}$ s$^{-1}$ arcsec$^{-2}$ ($3\times10^{-9}$ cm$^{-2}$ s$^{-1}$ arcsec$^{-2}$ at $E\sim0.3-2.0$ keV). The extended surface brightness limit is substantially dimmer ($\times1/500$) than the reference point source sensitivity.

The use of large spatial binning is not needed for the brighter regions of emission. Adaptive binning methods such as CSMOOTH \citep{ebeling+2006mnras368_65} or Voronoi binning \citep{cappellari+2003mnras342_345, diehl+2006mnras368_497} take advantage of the anisotropies of the emission in the detector and bin the low surface brightness regions (typically the outskirts or halos) more aggressively than for brighter sections of the image (typically the core of the galaxy). These methods, including Voronoi binning (used by \SAUNAS), have been extensively used in the literature to analyze extended low surface brightness emission Chandra and XMM-Newton observations \citep[see ][and references therein]{ebeling+2007apj661_33,gonzalezmartin+2009aap506_1107,broos+2010apj714_1582, ebeling+2010mnras407_83, xue+2011apj195_10, hodgeskluck+2012apj746_167, wang+2024apj962_188}. While \citet[][\SAUNAS\, I]{borlaff+2024apj967_169} provides extensive quality-checks (see Sec.\,2.3 in the original paper), we triple-check the results presented in Sec.\,\ref{subsec:results_xray_ima} by presenting two additional tests to assess the detection of extended emission around the core of NGC\,5084 (see Appendix \ref{Appendix:Xray_noPSFdeco_test} and \ref{Appendix:Xray_novoro_test}).

\subsection{Detection of X-ray extended emission without PSF deconvolution or Voronoi binning}
\label{Appendix:Xray_noPSFdeco_test}

To test if the significance of the detected extended X-ray emission is independent of the processing methodology applied by \SAUNAS\ (PSF deconvolution, Voronoi binning), this section applies an alternative methodology widely used in extragalactic X-ray astronomy to detect extended sources: compare the radial emission profile of the target with that of the PSF of the instrument \citep{fabbiano+2017apj842_4,fabbiano+2018apj855_131, jones+2020apj891_133,ma+2020apj900_164,ma+2023apj948_61}. If a significant excess flux is detected above those predicted by the scaled PSF, then extended emission must be present in the source.

We simulate the \Chandra/ACIS PSF using the \texttt{MARX} software following the procedure described in the \ciao\ documentation\footnote{\ciao/\texttt{MARX}: \url{https://cxc.cfa.harvard.edu/ciao/threads/marx_sim/}}. The PSF takes into account the exact location of the source in the focal plane of ACIS, and the spectral energy distribution of the source as well. The PSF generation process is also described in steps 4--6 of Sec.\,2.2.1 in \citet{borlaff+2024apj967_169}. To ensure that the PSF profile is well-sampled at high radii (associated with the outskirts of the galaxy, the region of interest), the generated PSF contains $3\times10^6$ events generated through multiple ray-tracing simulations (more than two orders of magnitude higher than all the events registered in the \Chandra/ACIS observations of NGC\,5084). This approach ensures that any potential excess of source counts compared with those of the PSF is not a false positive.

To generate the galaxy profile, all point sources (potentially associated with XRBs or background quasars) are removed similar to \SAUNAS's approach, with the exception of the core of the galaxy which, given its role as the main contributor to the scattered light emission, is left intact. Finally, the PSF central surface brightness profile is scaled to the central surface brightness of the observed galaxy, comparing their radial variation of surface brightness.

\begin{figure*}[]
\begin{center}
\begin{overpic}[trim={60 0 60 0}, clip, width=\textwidth]{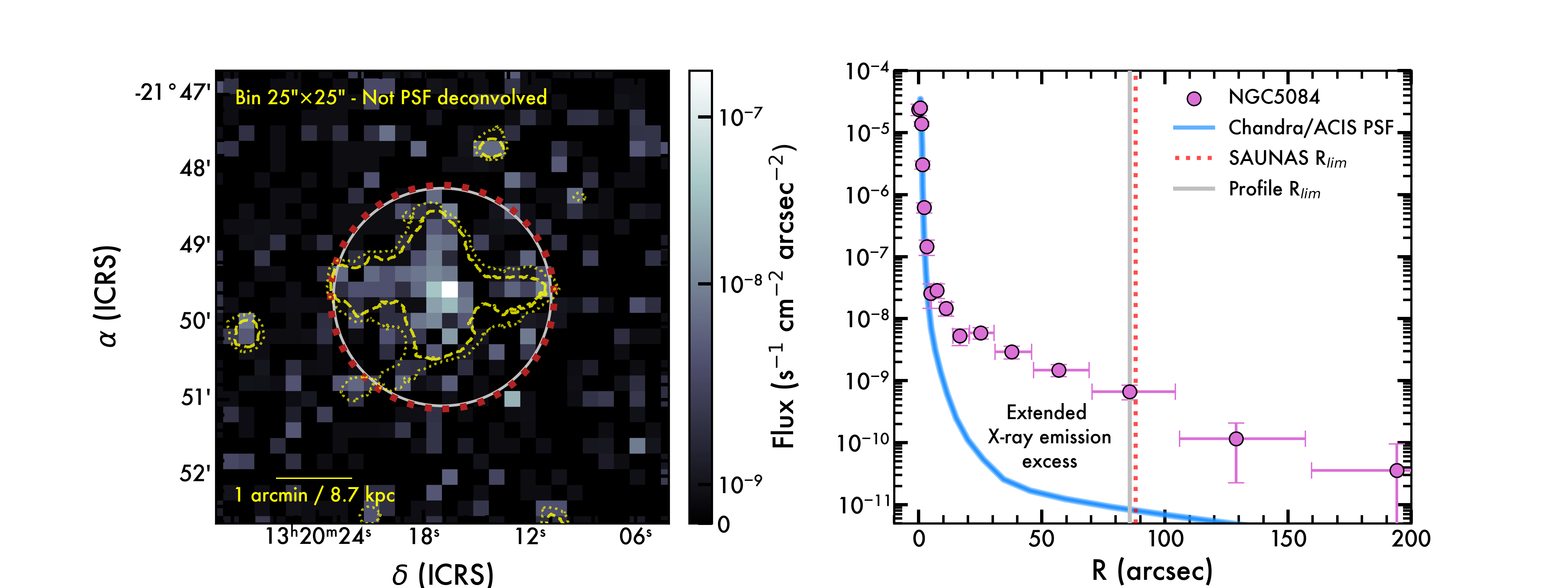}
\end{overpic}

\caption{Detection of the X-ray extended emission around NGC\,5084 without PSF deconvolution or adaptive smoothing. \emph{Left:} \Chandra/ACIS X-ray flux ($0.3-2.0$ keV) map rebinned to $12.2\times12.2$ arcsec$^2$ ($25\times25$ pixels) for visualization purposes. No PSF deconvolution or adaptive smoothing was applied (i.e., Voronoi binning). \emph{Yellow contours:} [$2$,$3$]$\sigma$ (dotted, dashed respectively) detection limits as estimated by \SAUNAS\ (see Sec.\,\ref{subsec:results_xray_ima}, for reference. \emph{Right:} Surface brightness profile of the non-PSF deconvolved, non-adaptively smoothed flux map (purple circles) and scaled \Chandra/ACIS PSF (blue solid line). \emph{Red dashed circle and vertical line:} \SAUNAS\ $3\sigma$ radial detection limit. \emph{Silver solid circle and vertical line:} Surface brightness profile radial detection limit of the non-PSF deconvolved, non-adaptive smoothed flux map. Sky background subtraction was applied to the flux map and surface brightness profile. We remark that the spatial binning of the flux map in the left panel of Fig.\,\ref{fig:NGC5084_psf_profile} is only for visualization purposes, and the surface brightness profile was generated based on the full resolution (0.492") dataset. See legend for details.}
\label{fig:NGC5084_psf_profile}
\end{center}
\end{figure*}

The results are shown in Fig.\,\ref{fig:NGC5084_psf_profile}. The left panel shows the broadband ($0.3-2.0$ keV) \Chandra/ACIS X-ray flux map as generated by the \ciao\ pipeline, demonstrating signs of potential extended emission. No PSF deconvolution or adaptive smoothing was applied in this image. In the right panel, the surface brightness profile of the broadband X-ray flux is compared to the scaled PSF of the \Chandra/ACIS observations at the position of NGC\,5084. The \Chandra/ACIS PSF at the observational setup of NGC\,5084 has a FWHM of 1.1".

Compared to the PSF and the background, we detect a $>>3\sigma$ excess of X-ray emission up to a radial distance of $\sim90$ arcsec from the core in the observations of NGC\,5084. Between 25-90 arcsec, the PSF scattered light is two orders of magnitude dimmer than the detected extended emission of NGC\,5084. The radial limit of the profile is compatible with the $3\sigma$ radial limit of the emission detected by \SAUNAS\ (see Sec.\,\ref{subsec:results_xray_ima}).

The test presented in this section demonstrates three important points:

\begin{enumerate}
    \item The extended emission of NGC\,5084 is not caused by PSF contamination from the strong core emission or by a Voronoi binning artifact.

    \item The statistical significance of the extended X-ray emission is independent of the PSF deconvolution and spatial binning applied.

    \item The total extension of NGC\,5084's broad-band emission in the \SAUNAS-generated maps is compatible with the $3\sigma$ limit measured using a completely independent methodology (X-ray surface brightness profile limit).
\end{enumerate}

\subsection{Independent detection of NGC\,5084 X-ray lobes without Voronoi binning}
\label{Appendix:Xray_novoro_test}

\begin{figure*}[]
\begin{center}
\begin{overpic}[trim={60 0 60 0}, clip, width=\textwidth]{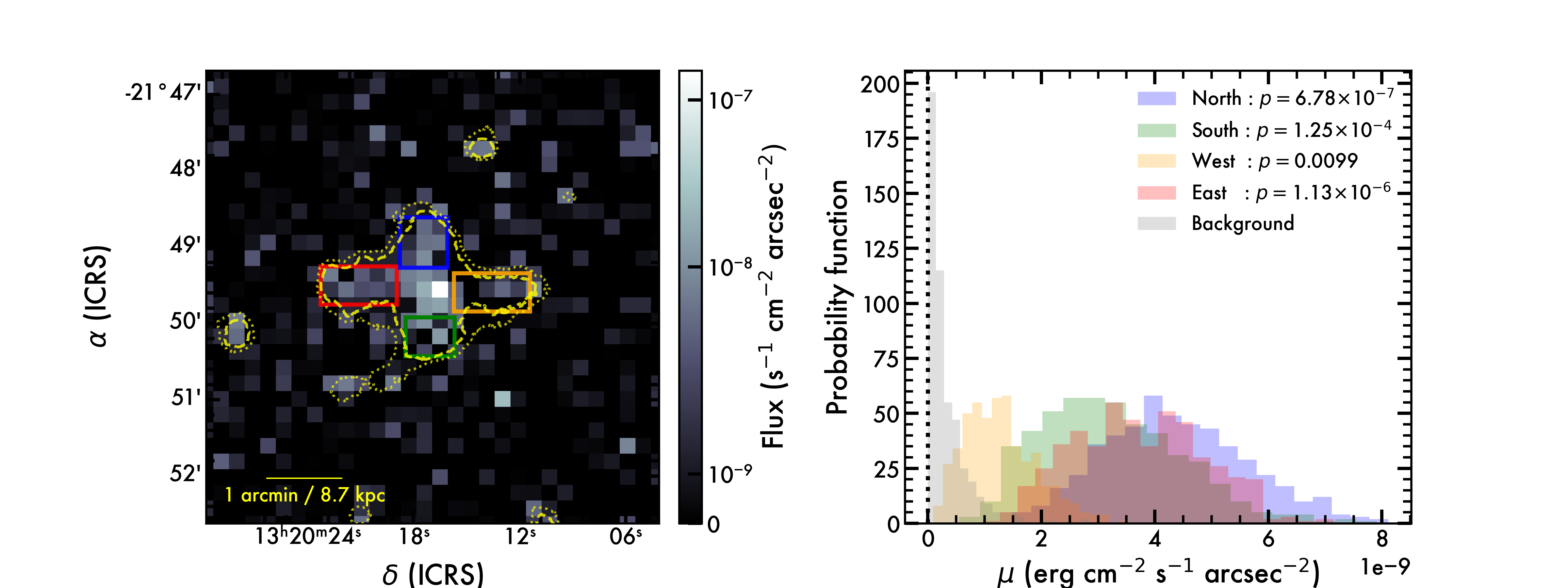}
\put(55,175){\color{yellow} {\textsf{Bin 25"$\times$25" - PSF deconvolved}}}
\end{overpic}
\caption{Detection of the X-ray extended lobes around NGC\,5084 without Voronoi binning. \emph{Left:} \Chandra/ACIS X-ray flux ($0.3-2.0$ keV) map rebinned to $12.2\times12.2$ arcsec ($25\times25$ pixels) for visualization purposes. PSF deconvolution was applied \citep[see Sec.\,\ref{sec:methods}, and ][]{borlaff+2024apj967_169} \emph{Color rectangles:} Fixed apertures to measure X-ray emission in the North (blue), West (red), South (green), and East (orange) lobes. \emph{Yellow contours:} [$2$,$3$]$\sigma$ (dotted, dashed respectively) detection limits as estimated by \SAUNAS\ (see Sec.\,\ref{subsec:results_xray_ima}, for reference. \emph{Right:} Color coded histograms represent the event probability distributions for the background (grey) and the four lobe apertures. The $p$-values for the null hypothesis that the flux distributions in the lobe apertures are compatible with the background are represented in the legend.}
\label{fig:NGC5084_Aperture_Histograms}
\end{center}
\end{figure*}

In this section we assess the statistical significance of the detection of the four X-ray emission lobes reported in Sec.\,\ref{subsec:results_xray_ima}, using the PSF deconvolved observations but without employing Voronoi binning. First, four simple box regions are defined, based on the contours detected in Sec.\,\ref{subsec:results_xray_ima} to isolate the emission from the brighter galactic core. The region parameters are listed in Table \ref{tab:Xray_flux} and represented in the left panel of Fig.\,\ref{fig:NGC5084_Aperture_Histograms} (see legend in the right panel). Second, a Poisson means test \citep[$E$-test, ][]{KRISHNAMOORTHY200423} evaluates the null hypothesis that the difference between the observed and background emission is statistically zero. We repeat the test 500 times using Monte Carlo and Bootstrapping simulations in order to obtain the probability distributions for the surface brightness represented in the right panel of Fig.\,\ref{fig:NGC5084_Aperture_Histograms}.

The analysis indicates that the surface brightness X-ray emission in the apertures is significantly higher than that of the background. The null hypothesis that the surface brightness distributions of lobe emission and the background are sampled from a common parent population is rejected at a p-value of $p<0.05$ in all four lobe regions. In other words, this analysis provides strong statistical support for the existence of the lobe emission: north ($p=6.8\times10^{-7}$), east ($p=1.1\times10^{-6}$), south ($p=1.3\times10^{-4}$) and west ($p\sim0.01$). The fluxes integrated over the different apertures are tabulated in Table \ref{tab:Xray_flux}. This result supports and verifies the findings described in Sec.\,\ref{subsec:results_xray_ima} and Appendix \ref{Appendix:Xray_noPSFdeco_test}. We conclude that the extended X-ray emission detected by \SAUNAS\ around NGC\,5084 is (1) statistically significant; (2) independent of the PSF deconvolution process; and  (3) independent of the Voronoi binning methodology applied.

\section{\Chandra\ X-ray images in sub-bands}
\label{Appendix:Xray_subbands}

\begin{figure*}[t!]
\begin{center}
 \begin{overpic}[trim={75 38 0 40}, clip, width=\textwidth]{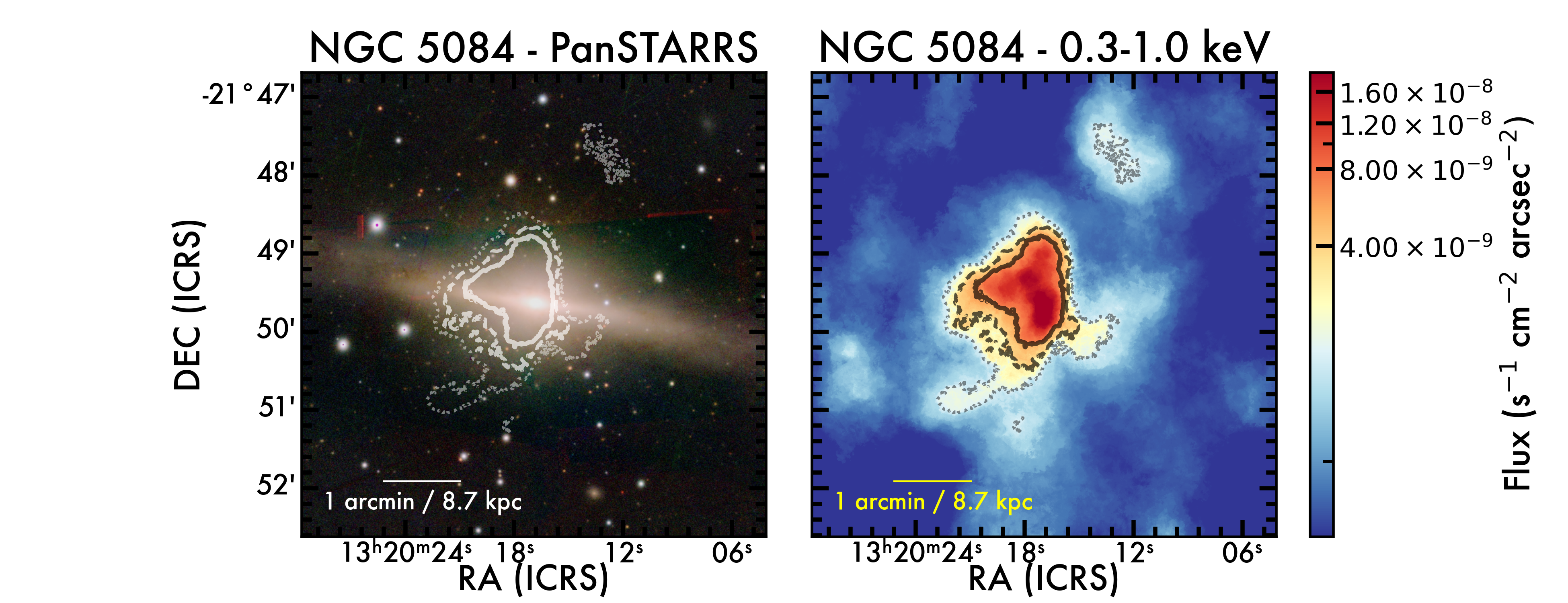}
\put(70,152.0){\large \color{yellow} {\textsf{0.3-1.0 keV}}}
\end{overpic}

\begin{overpic}[trim={75 38 0 40}, clip, width=\textwidth]{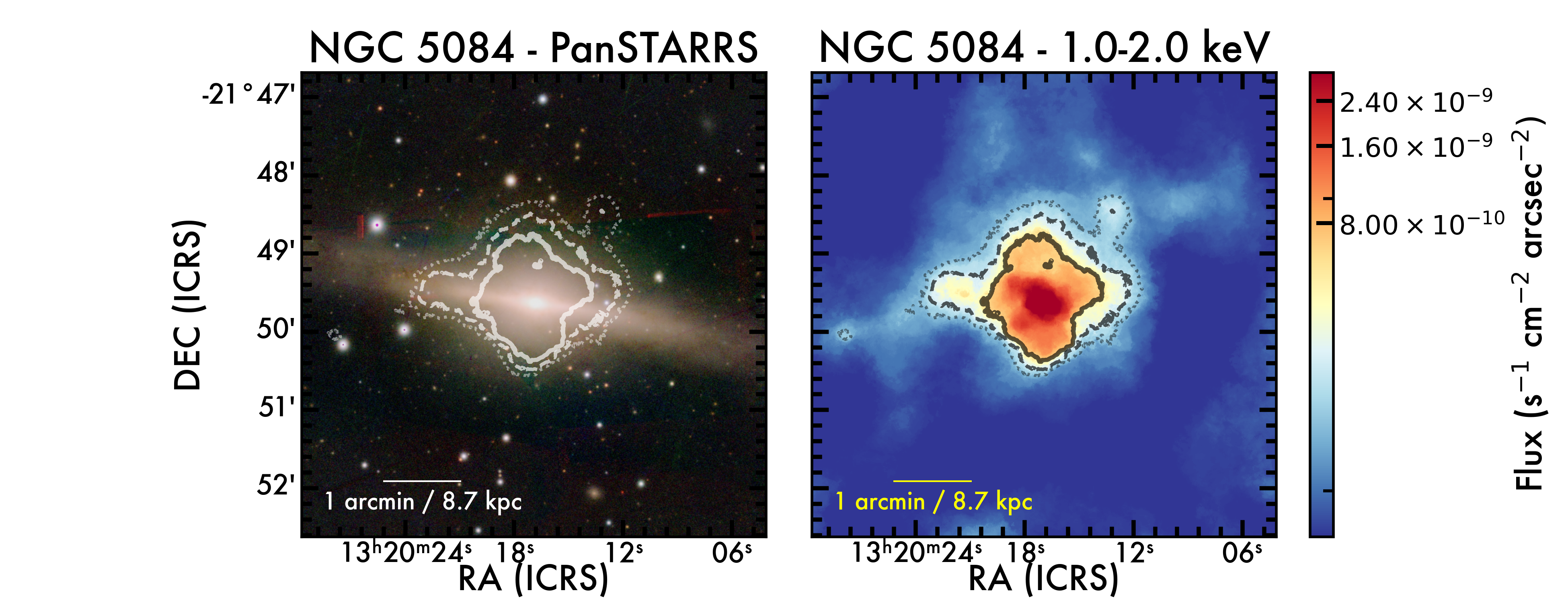}
\put(70,152.0){\large \color{yellow} {\textsf{1.0-2.0 keV}}}
\end{overpic}

\begin{overpic}[trim={75 0 0 38}, clip, width=\textwidth]{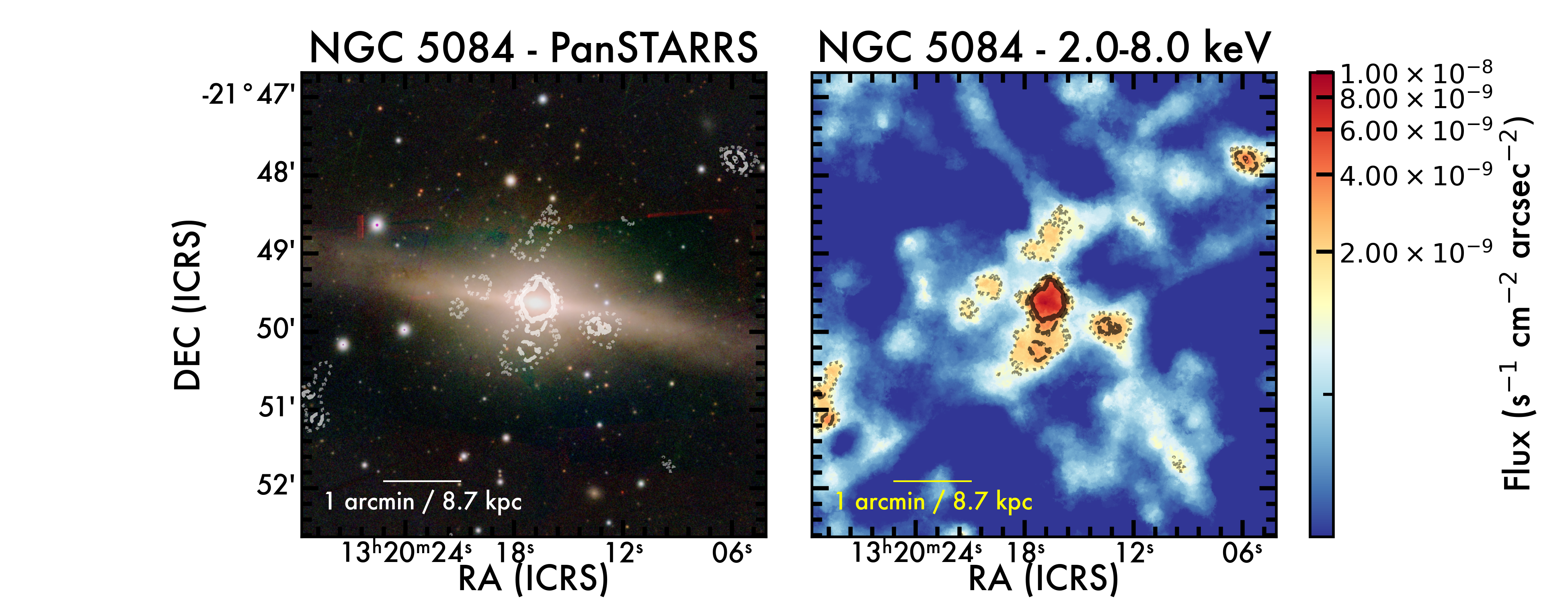}
\put(70,173.0){\large \color{yellow} {\textsf{2.0-8.0 keV}}}
\end{overpic}

\caption{Diffuse X-ray emission of NGC\,5084 as detected with \SAUNAS/\Chandra\ in 0.3--1.0 keV (\emph{top}), 1.0--2.0 keV (\emph{center}), 2.0--8.0 keV (\emph{bottom}). \emph{Left:} Optical $gri$ color Pan-STARRS image \citep{chambers+2016arXiv1612.05560}. \emph{Right:} \SAUNAS\ map of the diffuse X-ray emission, corrected for PSF, point-sources, and sky-background. Dashed and solid contours represent the $2\sigma$ and the $3\sigma$ detection levels of X-ray emission, represented in white (left panel) and black (right panel) for contrast.}
\label{fig:NGC5084_per_band}
\end{center}
\end{figure*}

Figure \ref{fig:NGC5084_per_band} shows the X-ray mosaics in the individual bands (\emph{soft:} 0.3-1.0 keV, \emph{medium:} 1.0-2.0 keV, \emph{hard:} 2.0 - 8.0 keV) processed with \texttt{SAUNAS I} \citep{borlaff+2024apj967_169}.

\section{\emph{Hubble} Space Telescope multiband imaging}
\label{Appendix:Hubble_subbands}

\begin{figure*}[]
\begin{center}
\includegraphics[trim={0 25 60 0}, clip, height=8.85cm]{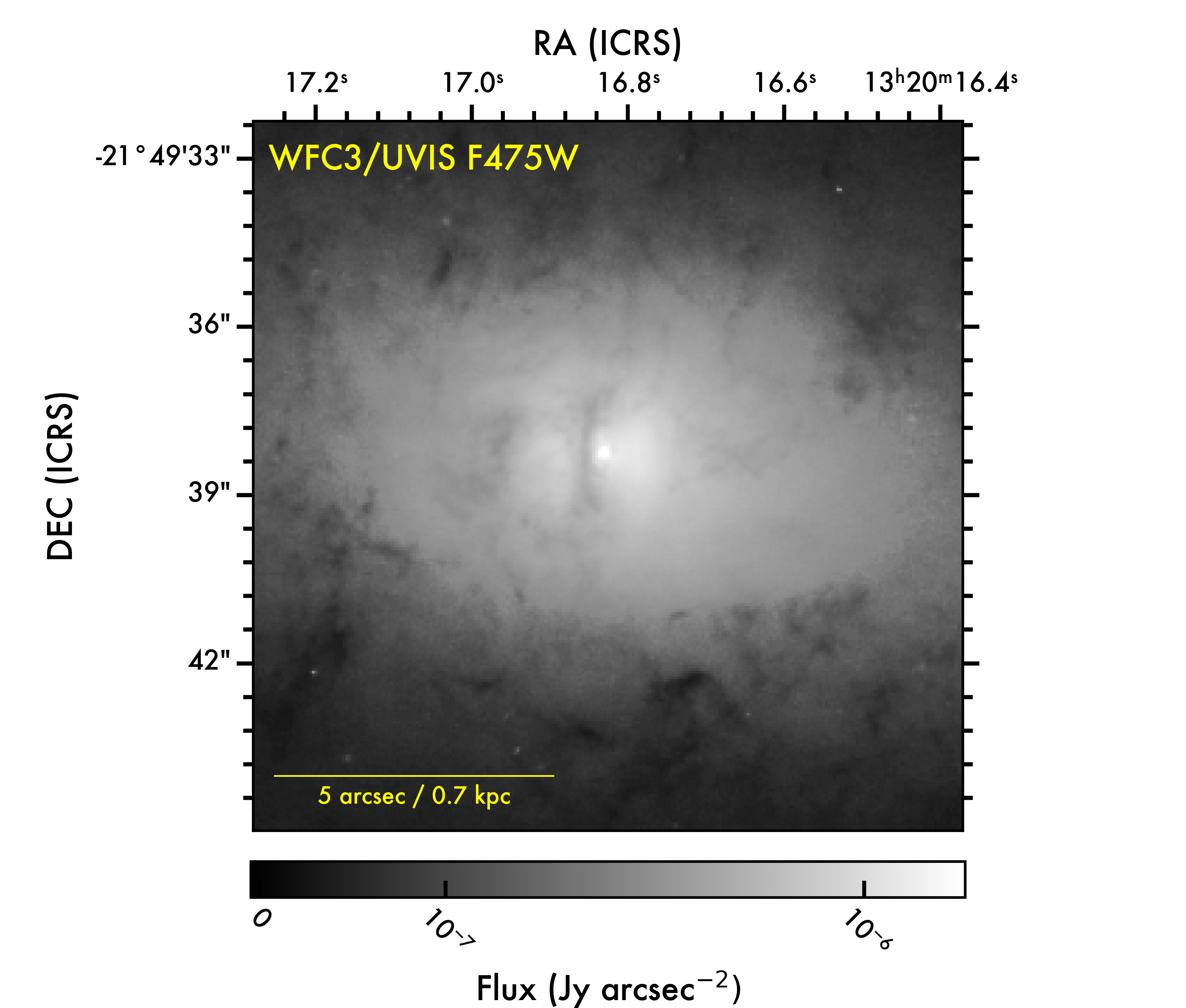}
\includegraphics[trim={140 25 60 0}, clip, height=8.85cm]{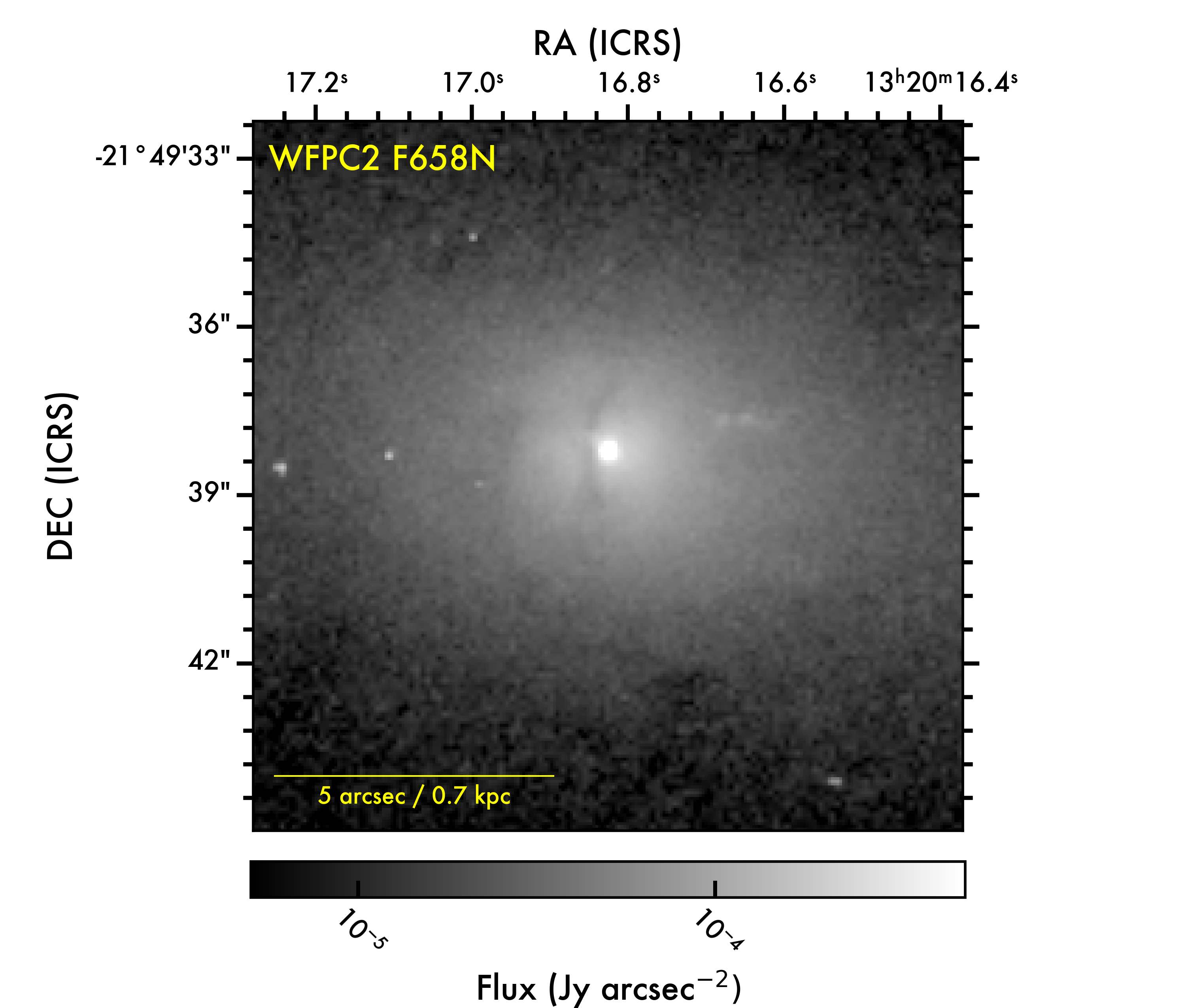}

\includegraphics[trim={0 0 60 57}, clip, height=8.4cm]{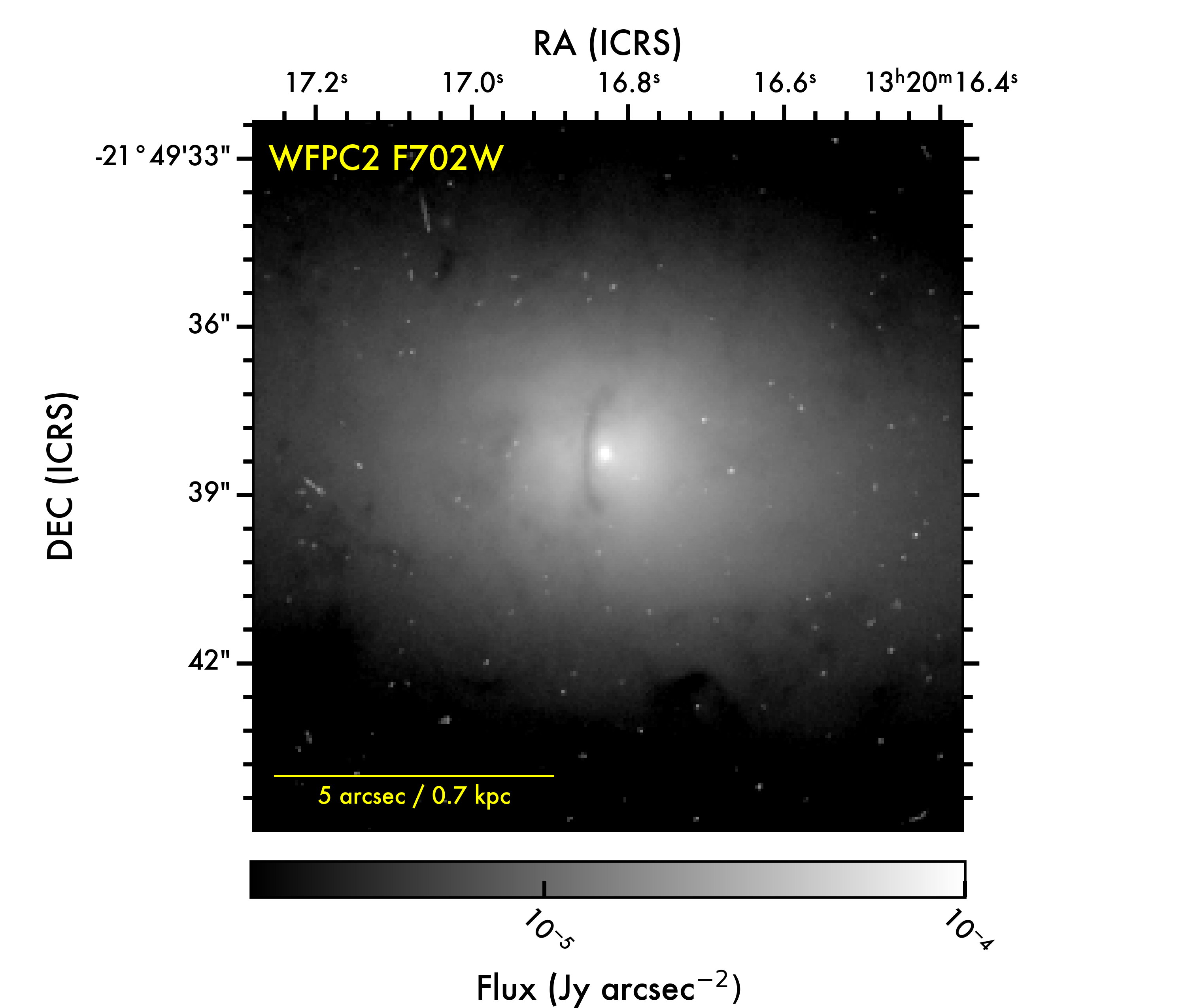}
\includegraphics[trim={140 0 60 57}, clip, height=8.4cm]{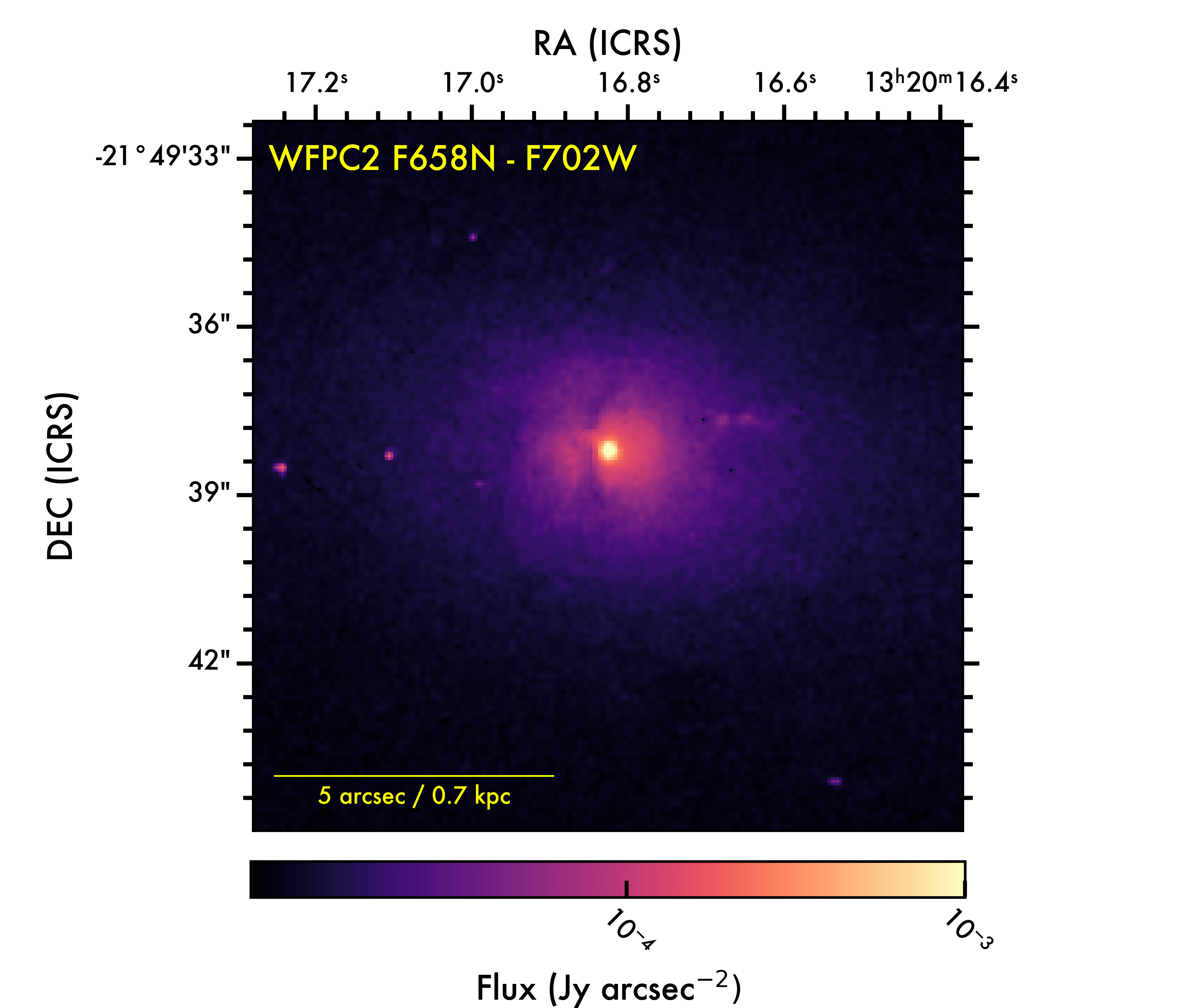}

\caption{Morphology of the core of NGC\,5084 obtained with \emph{Hubble}. \emph{Top left:} WFC3/UVIS F475W, \emph{Top right:} WFPC2 F658N, \emph{Bottom left:} WFPC2 F702W, \emph{Bottom right:} F658N - F702W (proxy of the [NII, $\lambda$6548] + $H\alpha$ emission). The area shown is $12.6\times12.6$~arcsec$^2$ ($1.82\times1.82$~kpc$^2$ at a distance of 29.91~Mpc). See the scales and colorbars for reference.}
\label{fig:NGC5084_multiband}
\end{center}
\end{figure*}

Figure \ref{fig:NGC5084_multiband} shows the emission in each individual HST band analyzed in Sec.\,\ref{subsec:results_hst}, including the F658N - F702W color image.

\label{Appendix:OpticalSpectra}


\bibliography{main}\bibliographystyle{aasjournal}



\end{document}